\documentclass[reprint,amsmath,amssymb,aps,floatfix,superscriptaddress,showpacs,prl]{revtex4-1}
\usepackage{graphicx}%
\usepackage{dcolumn}%
\usepackage{bm}%
\usepackage{color}
\usepackage{subfigure}
\usepackage{url}

\usepackage{comment}

\usepackage[colorlinks]{hyperref}
\hypersetup{bookmarks, colorlinks=true, citecolor=cyan, filecolor=blue ,linkcolor=red, urlcolor=blue}

\begin{document}
{
\title{Precision spectral measurements of Chromium and Titanium from 10 to 250 GeV/\textit{n} and sub-Iron to Iron ratio with the Calorimetric Electron Telescope on the ISS}

\author{O.~Adriani}
\affiliation{Department of Physics, University of Florence, Via Sansone, 1 - 50019, Sesto Fiorentino, Italy}
\affiliation{INFN Sezione di Firenze, Via Sansone, 1 - 50019, Sesto Fiorentino, Italy}
\author{Y.~Akaike}
\email{yakaike@aoni.waseda.jp}
\affiliation{Waseda Research Institute for Science and Engineering, Waseda University, 17 Kikuicho,  Shinjuku, Tokyo 162-0044, Japan}
\affiliation{JEM Utilization Center, Human Spaceflight Technology Directorate, Japan Aerospace Exploration Agency, 2-1-1 Sengen, Tsukuba, Ibaraki 305-8505, Japan}
\author{K.~Asano}
\affiliation{Institute for Cosmic Ray Research, The University of Tokyo, 5-1-5 Kashiwa-no-Ha, Kashiwa, Chiba 277-8582, Japan}
\author{Y.~Asaoka}
\affiliation{Institute for Cosmic Ray Research, The University of Tokyo, 5-1-5 Kashiwa-no-Ha, Kashiwa, Chiba 277-8582, Japan}
\author{E.~Berti} 
\affiliation{INFN Sezione di Firenze, Via Sansone, 1 - 50019, Sesto Fiorentino, Italy}
\affiliation{Institute of Applied Physics (IFAC),  National Research Council (CNR), Via Madonna del Piano, 10, 50019, Sesto Fiorentino, Italy}
\author{P.~Betti}
\affiliation{INFN Sezione di Firenze, Via Sansone, 1 - 50019, Sesto Fiorentino, Italy}
\affiliation{Institute of Applied Physics (IFAC),  National Research Council (CNR), Via Madonna del Piano, 10, 50019, Sesto Fiorentino, Italy}
\author{G.~Bigongiari}
\affiliation{Department of Physical Sciences, Earth and Environment, University of Siena, via Roma 56, 53100 Siena, Italy}
\affiliation{INFN Sezione di Pisa, Polo Fibonacci, Largo B. Pontecorvo, 3 - 56127 Pisa, Italy}
\author{W.R.~Binns}
\affiliation{Department of Physics and McDonnell Center for the Space Sciences, Washington University, One Brookings Drive, St. Louis, Missouri 63130-4899, USA}
\author{M.~Bongi}
\affiliation{Department of Physics, University of Florence, Via Sansone, 1 - 50019, Sesto Fiorentino, Italy}
\affiliation{INFN Sezione di Firenze, Via Sansone, 1 - 50019, Sesto Fiorentino, Italy}
\author{P.~Brogi}
\affiliation{Department of Physical Sciences, Earth and Environment, University of Siena, via Roma 56, 53100 Siena, Italy}
\affiliation{INFN Sezione di Pisa, Polo Fibonacci, Largo B. Pontecorvo, 3 - 56127 Pisa, Italy}
\author{A.~Bruno}
\affiliation{Heliospheric Physics Laboratory, NASA/GSFC, Greenbelt, Maryland 20771, USA}
\author{N.~Cannady}
\affiliation{Astroparticle Physics Laboratory, NASA/GSFC, Greenbelt, Maryland 20771, USA}
\author{G.~Castellini}
\affiliation{Institute of Applied Physics (IFAC),  National Research Council (CNR), Via Madonna del Piano, 10, 50019, Sesto Fiorentino, Italy}
\author{C.~Checchia}
\email[]{caterina.checchia2@unisi.it}
\affiliation{Department of Physical Sciences, Earth and Environment, University of Siena, via Roma 56, 53100 Siena, Italy}
\affiliation{INFN Sezione di Pisa, Polo Fibonacci, Largo B. Pontecorvo, 3 - 56127 Pisa, Italy}
\author{M.L.~Cherry}
\affiliation{Department of Physics and Astronomy, Louisiana State University, 202 Nicholson Hall, Baton Rouge, Louisiana 70803, USA}
\author{G.~Collazuol}
\affiliation{Department of Physics and Astronomy, University of Padova, Via Marzolo, 8, 35131 Padova, Italy}
\affiliation{INFN Sezione di Padova, Via Marzolo, 8, 35131 Padova, Italy} 
\author{G.A.~de~Nolfo}
\affiliation{Heliospheric Physics Laboratory, NASA/GSFC, Greenbelt, Maryland 20771, USA}
\author{K.~Ebisawa}
\affiliation{Institute of Space and Astronautical Science, Japan Aerospace Exploration Agency, 3-1-1 Yoshinodai, Chuo, Sagamihara, Kanagawa 252-5210, Japan}
\author{A.~W.~Ficklin}
\affiliation{Department of Physics and Astronomy, Louisiana State University, 202 Nicholson Hall, Baton Rouge, Louisiana 70803, USA}
\author{H.~Fuke}
\affiliation{Institute of Space and Astronautical Science, Japan Aerospace Exploration Agency, 3-1-1 Yoshinodai, Chuo, Sagamihara, Kanagawa 252-5210, Japan}
\author{S.~Gonzi}
\affiliation{Department of Physics, University of Florence, Via Sansone, 1 - 50019, Sesto Fiorentino, Italy}
\affiliation{INFN Sezione di Firenze, Via Sansone, 1 - 50019, Sesto Fiorentino, Italy}
\affiliation{Institute of Applied Physics (IFAC),  National Research Council (CNR), Via Madonna del Piano, 10, 50019, Sesto Fiorentino, Italy}
\author{T.G.~Guzik}
\affiliation{Department of Physics and Astronomy, Louisiana State University, 202 Nicholson Hall, Baton Rouge, Louisiana 70803, USA}
\author{T.~Hams}
\affiliation{Center for Space Sciences and Technology, University of Maryland, Baltimore County, 1000 Hilltop Circle, Baltimore, Maryland 21250, USA}
\author{K.~Hibino}
\affiliation{Kanagawa University, 3-27-1 Rokkakubashi, Kanagawa, Yokohama, Kanagawa 221-8686, Japan}
\author{M.~Ichimura}
\affiliation{Faculty of Science and Technology, Graduate School of Science and Technology, Hirosaki University, 3, Bunkyo, Hirosaki, Aomori 036-8561, Japan}

\author{M.H.~Israel}
\affiliation{Department of Physics and McDonnell Center for the Space Sciences, Washington University, One Brookings Drive, St. Louis, Missouri 63130-4899, USA}
\author{K.~Kasahara}
\affiliation{Department of Electronic Information Systems, Shibaura Institute of Technology, 307 Fukasaku, Minuma, Saitama 337-8570, Japan}
\author{J.~Kataoka}
\affiliation{School of Advanced Science and Engineering, Waseda University, 3-4-1 Okubo, Shinjuku, Tokyo 169-8555, Japan}
\author{R.~Kataoka}
\affiliation{National Institute of Polar Research, 10-3, Midori-cho, Tachikawa, Tokyo 190-8518, Japan}
\author{Y.~Katayose}
\affiliation{Faculty of Engineering, Division of Intelligent Systems Engineering, Yokohama National University, 79-5 Tokiwadai, Hodogaya, Yokohama 240-8501, Japan}
\author{C.~Kato}
\affiliation{Faculty of Science, Shinshu University, 3-1-1 Asahi, Matsumoto, Nagano 390-8621, Japan}
\author{N.~Kawanaka}
\affiliation{Department of Physics, Graduate School of Science, Tokyo Metropolitan University, 1-1 Minamii-Osawa, Hachioji, Tokyo 192-0397, Japan }
\affiliation{National Astronomical Observatory of Japan,  2-21-1 Osawa,  Mitaka, Tokyo 181-8588, Japan}
\author{Y.~Kawakubo}
\affiliation{ Department of Physical Sciences, College of Science and Engineering, Aoyama Gakuin University,  5-10-1 Fuchinobe, Chuo, Sagamihara, Kanagawa 252-5258, Japan}
\author{K.~Kobayashi}
\affiliation{Waseda Research Institute for Science and Engineering, Waseda University, 17 Kikuicho,  Shinjuku, Tokyo 162-0044, Japan}
\affiliation{JEM Utilization Center, Human Spaceflight Technology Directorate, Japan Aerospace Exploration Agency, 2-1-1 Sengen, Tsukuba, Ibaraki 305-8505, Japan}
\author{K.~Kohri} 
\affiliation{National Astronomical Observatory of Japan,  2-21-1 Osawa,  Mitaka, Tokyo 181-8588, Japan}
\affiliation{Institute of Particle and Nuclear Studies, High Energy Accelerator Research Organization, 1-1 Oho, Tsukuba, Ibaraki 305-0801, Japan} 
\author{H.S.~Krawczynski}
\affiliation{Department of Physics and McDonnell Center for the Space Sciences, Washington University, One Brookings Drive, St. Louis, Missouri 63130-4899, USA}
\author{J.F.~Krizmanic}
\affiliation{Astroparticle Physics Laboratory, NASA/GSFC, Greenbelt, Maryland 20771, USA}
\author{P.~Maestro}
\affiliation{Department of Physical Sciences, Earth and Environment, University of Siena, via Roma 56, 53100 Siena, Italy}
\affiliation{INFN Sezione di Pisa, Polo Fibonacci, Largo B. Pontecorvo, 3 - 56127 Pisa, Italy}
\author{P.S.~Marrocchesi}
\affiliation{Department of Physical Sciences, Earth and Environment, University of Siena, via Roma 56, 53100 Siena, Italy}
\affiliation{INFN Sezione di Pisa, Polo Fibonacci, Largo B. Pontecorvo, 3 - 56127 Pisa, Italy}
\author{M.~Mattiazzi}
\affiliation{Department of Physics and Astronomy, University of Padova, Via Marzolo, 8, 35131 Padova, Italy}
\affiliation{INFN Sezione di Padova, Via Marzolo, 8, 35131 Padova, Italy} 
\author{A.M.~Messineo}
\affiliation{INFN Sezione di Pisa, Polo Fibonacci, Largo B. Pontecorvo, 3 - 56127 Pisa, Italy}
\affiliation{University of Pisa, Polo Fibonacci, Largo B. Pontecorvo, 3 - 56127 Pisa, Italy}
\author{J.W.~Mitchell}
\affiliation{Astroparticle Physics Laboratory, NASA/GSFC, Greenbelt, Maryland 20771, USA}
\author{S.~Miyake}
\affiliation{Department of Electrical and Computer Engineering, National Institute of Technology (KOSEN), Gifu College, 2236-2 Kamimakuwa, Motosu-city, Gifu 501-0495, Japan}
\author{A.A.~Moiseev}
\affiliation{Astroparticle Physics Laboratory, NASA/GSFC, Greenbelt, Maryland 20771, USA}
\affiliation{Center for Research and Exploration in Space Sciences and Technology, NASA/GSFC, Greenbelt, Maryland 20771, USA}
\affiliation{Department of Astronomy, University of Maryland, College Park, Maryland 20742, USA}\author{M.~Mori}
\affiliation{Department of Physical Sciences, College of Science and Engineering, Ritsumeikan University, Shiga 525-8577, Japan}
\author{N.~Mori}
\affiliation{INFN Sezione di Firenze, Via Sansone, 1 - 50019, Sesto Fiorentino, Italy}
\author{H.M.~Motz}
\affiliation{Faculty of Science and Engineering, Global Center for Science and Engineering, Waseda University, 3-4-1 Okubo, Shinjuku, Tokyo 169-8555, Japan}
\author{K.~Munakata}
\affiliation{Faculty of Science, Shinshu University, 3-1-1 Asahi, Matsumoto, Nagano 390-8621, Japan}
\author{S.~Nakahira}
\affiliation{Institute of Space and Astronautical Science, Japan Aerospace Exploration Agency, 3-1-1 Yoshinodai, Chuo, Sagamihara, Kanagawa 252-5210, Japan}
\author{J.~Nishimura}
\affiliation{Institute of Space and Astronautical Science, Japan Aerospace Exploration Agency, 3-1-1 Yoshinodai, Chuo, Sagamihara, Kanagawa 252-5210, Japan}

\author{M.~Negro}
\affiliation{Department of Physics and Astronomy, Louisiana State University, 202 Nicholson Hall, Baton Rouge, Louisiana 70803, USA}
\author{S.~Okuno}
\affiliation{Kanagawa University, 3-27-1 Rokkakubashi, Kanagawa, Yokohama, Kanagawa 221-8686, Japan}
\author{J.F.~Ormes}
\affiliation{Department of Physics and Astronomy, University of Denver, Physics Building, Room 211, 2112 East Wesley Avenue, Denver, Colorado 80208-6900, USA}
\author{S.~Ozawa}
\affiliation{Quantum ICT Advanced Development Center, National Institute of Information and Communications Technology, 4-2-1 Nukui-Kitamachi, Koganei, Tokyo 184-8795, Japan}
\author{L.~Pacini}
\affiliation{INFN Sezione di Firenze, Via Sansone, 1 - 50019, Sesto Fiorentino, Italy}
\affiliation{Institute of Applied Physics (IFAC),  National Research Council (CNR), Via Madonna del Piano, 10, 50019, Sesto Fiorentino, Italy}
\author{P.~Papini}
\affiliation{INFN Sezione di Firenze, Via Sansone, 1 - 50019, Sesto Fiorentino, Italy}
\author{B.F.~Rauch}
\affiliation{Department of Physics and McDonnell Center for the Space Sciences, Washington University, One Brookings Drive, St. Louis, Missouri 63130-4899, USA}
\author{S.B.~Ricciarini}
\affiliation{INFN Sezione di Firenze, Via Sansone, 1 - 50019, Sesto Fiorentino, Italy}
\affiliation{Institute of Applied Physics (IFAC),  National Research Council (CNR), Via Madonna del Piano, 10, 50019, Sesto Fiorentino, Italy}
\author{K.~Sakai}
\affiliation{Kavli Institute for Cosmological Physics, The University of Chicago,  5640 South Ellis Avenue, Chicago, IL 60637, USA}
\author{T.~Sakamoto}
\affiliation{ Department of Physical Sciences, College of Science and Engineering, Aoyama Gakuin University,  5-10-1 Fuchinobe, Chuo, Sagamihara, Kanagawa 252-5258, Japan}
\author{M.~Sasaki}
\affiliation{Astroparticle Physics Laboratory, NASA/GSFC, Greenbelt, Maryland 20771, USA}
\affiliation{Center for Research and Exploration in Space Sciences and Technology, NASA/GSFC, Greenbelt, Maryland 20771, USA}
\affiliation{Department of Astronomy, University of Maryland, College Park, Maryland 20742, USA}
\author{Y.~Shimizu}
\affiliation{Kanagawa University, 3-27-1 Rokkakubashi, Kanagawa, Yokohama, Kanagawa 221-8686, Japan}
\author{A.~Shiomi}
\affiliation{College of Industrial Technology, Nihon University, 1-2-1 Izumi, Narashino, Chiba 275-8575, Japan}
\author{P.~Spillantini}
\affiliation{Department of Physics, University of Florence, Via Sansone, 1 - 50019, Sesto Fiorentino, Italy}
\author{F.~Stolzi}
\email[]{francesco.stolzi@unisi.it}
\affiliation{Department of Physical Sciences, Earth and Environment, University of Siena, via Roma 56, 53100 Siena, Italy}
\affiliation{INFN Sezione di Pisa, Polo Fibonacci, Largo B. Pontecorvo, 3 - 56127 Pisa, Italy}
\author{S.~Sugita}
\affiliation{ Department of Physical Sciences, College of Science and Engineering, Aoyama Gakuin University,  5-10-1 Fuchinobe, Chuo, Sagamihara, Kanagawa 252-5258, Japan}
\author{A.~Sulaj} 
\affiliation{Department of Physical Sciences, Earth and Environment, University of Siena, via Roma 56, 53100 Siena, Italy}
\affiliation{INFN Sezione di Pisa, Polo Fibonacci, Largo B. Pontecorvo, 3 - 56127 Pisa, Italy}
\author{M.~Takita}
\affiliation{Institute for Cosmic Ray Research, The University of Tokyo, 5-1-5 Kashiwa-no-Ha, Kashiwa, Chiba 277-8582, Japan}
\author{T.~Tamura}
\affiliation{Kanagawa University, 3-27-1 Rokkakubashi, Kanagawa, Yokohama, Kanagawa 221-8686, Japan}
\author{T.~Terasawa}
\affiliation{Institute for Cosmic Ray Research, The University of Tokyo, 5-1-5 Kashiwa-no-Ha, Kashiwa, Chiba 277-8582, Japan}
\author{S.~Torii}
\affiliation{Waseda Research Institute for Science and Engineering, Waseda University, 17 Kikuicho,  Shinjuku, Tokyo 162-0044, Japan}
\author{Y.~Tsunesada}
\affiliation{Graduate School of Science, Osaka Metropolitan University, Sugimoto, Sumiyoshi, Osaka 558-8585, Japan }
\affiliation{Nambu Yoichiro Institute for Theoretical and Experimental Physics, Osaka Metropolitan University,  Sugimoto, Sumiyoshi, Osaka  558-8585, Japan}
\author{Y.~Uchihori}
\affiliation{National Institutes for Quantum and Radiation Science and Technology, 4-9-1 Anagawa, Inage, Chiba 263-8555, Japan}
\author{E.~Vannuccini}
\affiliation{INFN Sezione di Firenze, Via Sansone, 1 - 50019, Sesto Fiorentino, Italy}
\author{J.P.~Wefel}
\affiliation{Department of Physics and Astronomy, Louisiana State University, 202 Nicholson Hall, Baton Rouge, Louisiana 70803, USA}
\author{K.~Yamaoka}
\affiliation{Nagoya University, Furo, Chikusa, Nagoya 464-8601, Japan}
\author{S.~Yanagita}
\affiliation{College of Science, Ibaraki University, 2-1-1 Bunkyo, Mito, Ibaraki 310-8512, Japan}
\author{A.~Yoshida}
\affiliation{ Department of Physical Sciences, College of Science and Engineering, Aoyama Gakuin University,  5-10-1 Fuchinobe, Chuo, Sagamihara, Kanagawa 252-5258, Japan}
\author{K.~Yoshida}
\affiliation{Department of Electronic Information Systems, Shibaura Institute of Technology, 307 Fukasaku, Minuma, Saitama 337-8570, Japan}
\author{W.~V.~Zober}
\affiliation{Department of Physics and McDonnell Center for the Space Sciences, Washington University, One Brookings Drive, St. Louis, Missouri 63130-4899, USA}
\collaboration{CALET Collaboration}

\date{\today}

\begin{abstract}
  The Calorimetric Electron Telescope (CALET), in operation on the International Space Station since 2015, collected a large sample of cosmic-ray (CR) iron and sub-iron events over a wide energy interval. In this paper we report an update of our previous measurement of the iron flux and we present - for the first time - a high statistics measurement of the spectra of two sub-iron elements Cr and Ti in the energy interval from 10 to 250 GeV$/n$. The analyses are based on 8 years of data.
  Differently from older generations of cosmic-ray instruments which, in most cases, could not resolve individual sub-iron elements, CALET can identify each nuclear species from proton to nickel (and beyond) with a measurement of their electric charge.

 Thanks to the improvement in statistics and a more refined assessment of systematic uncertainties, the iron spectral shape is better resolved, at high energy, than in our previous paper and we report its flux ratio to chromium and titanium.
 
 The measured fluxes of Cr and Ti show energy dependences compatible with a single power law with spectral indices $ -2.74 \pm 0.06$ and  $ -2.88 \pm 0.06$, respectively.

\end{abstract}

\maketitle

\section{Introduction}

Cosmic-ray elements  lying just below iron in the periodic table (sub-iron),  like scandium (Sc), titanium (Ti), vanadium (V), chromium (Cr), and manganese (Mn), are thought to be mainly of secondary origin, i.e., produced by the spallation of heavier nuclei as they propagate in the interstellar medium, although for some a primary component from astrophysical sources cannot be excluded \cite{Boschini2021,HNE}.
As such, they provide  insight into cosmic ray propagation through the galaxy. In particular, the ratios of sub-iron elements to iron provide unique information on the average path length that cosmic rays travel before reaching Earth. 
The traditional “Leaky Box Model” \cite{BoxModel} interpretation of secondary-to-primary ratios as a function of energy is challenged by new theoretical approaches  \cite{Tomassetti, Aloisio2015, Johannesson,Cowsik, Bresci, Evoli2019,Cuoco,Cowsik2} driven by new measurements \cite{CALET2022BtoC, PAMELA2007, DAMPE-BtoC, AMS-Li-Be-B} with unmatched precision and data sets orders of magnitude larger than earlier experiments.
 A number of models have been proposed to provide a comprehensive description of secondary production and acceleration including, but not limited to: an inhomogeneous or energy-dependent diffusion coefficient  \cite{Tomassetti, Aloisio2015, Johannesson}; the production of a small fraction of secondaries by interactions of primary nuclei with matter  inside the acceleration region \cite{Cowsik, Bresci, Evoli2019}; the possible reacceleration of secondary particles during propagation \cite{Cuoco}; the presence of an energy independent term in the secondary-to-primary ratio \cite{Cowsik2}.\\
 \indent
Early direct measurements of sub-iron fluxes were carried out mainly by balloon-borne instruments followed by measurements in space.
 The best balloon data in this charge and energy region, from Simon, M. et al.~\cite{Simon1980} and Juliusson et al. (HEN)~\cite{HEN}  were severely limited by statistics.
 Satellite measurements were carried out by the Danish-French experiment \cite{HEAO} and the Heavy-Nuclei Experiment (HNE)~\cite{HNE}, both on the HEAO-3 satellite. HNE could measure the relative abundance to Fe of individual elements from  $_{18}$Ar to $_{23}$V, and $_{28}$Ni, from 10 GeV to several hundred GeV per amu, covering energies one order of magnitude larger than the Danish-French~experiment (hereafter HEAO3-C2).\\
\indent In the last two decades, direct measurements of charged cosmic rays  from space have achieved a level of unprecedented precision with long term observations of individual elements by AMS-02~\cite{AMS-Fe,AMS-CO,AMS-Li-Be-B,AMS-Ne-Mg-Si}, DAMPE~\cite{DAMPE-BtoC} and CALET~\cite{CALET-CO,CALET-IRON2021, CALET2022Ni, CALET2022BtoC,CALET-He}. However, no high-precision measurements of individual sub-iron fluxes have  yet been reported.\\
 {\indent}In this Letter we present the fluxes of Cr and Ti measured by CALET in the energy interval from 10 to 250~GeV/\textit{n} and their ratios to iron flux using the result of an updated analysis of iron based on a larger data set  from 10 GeV/\textit{n} to 1.6 TeV/\textit{n}  (see our previous publication \cite{CALET-IRON2021}). The new results suggest the presence of a spectral hardening for iron above 200~GeV/\textit{n}.

\section{CALET Instrument}
CALET is a  space-based instrument~\cite{CALET2021, CALET, CALET2, CALET3} optimized for the measurement of the electron+positron spectrum~\cite{CALET-ELE2017,CALET-ELE2018,CALET-ELE2023},
but also designed to study individual elements -- from proton to iron and above -- exploring particle energies up to the PeV scale, thanks to its large dynamic range, adequate calorimetric depth, accurate tracking, and excellent charge identification capabilities. 
CALET measures the particle's energy with the TASC (Total AbSorption Calorimeter), a thick lead-tungstate homogeneous calorimeter (27 radiation lengths, 1.2 proton interaction lengths) preceded by a thin (3 radiation lengths) pre-shower IMaging Calorimeter (IMC), both covering a very large dynamic range.  
Charge identification is carried out by a dedicated CHarge Detector (CHD), a two-layered hodoscope of plastic scintillator paddles placed on top of CALET.  The CHD can resolve individual elements from Z~=~1 to Z~=~40 with excellent charge resolution.  The IMC, with 16 layers of 1mm$^{2}$ square scintillating fibers (read out individually), provides tracking and an independent charge measurement, via multiple \textit{dE/dx} from fibers,  up to the onset of saturation which occurs for elements above  $_{14}$Si. 
Details on the instrument layout and the trigger system can be found in the Supplemental Material (SM) of Ref.~\cite{CALET-ELE2017}. 
CALET was launched on August 19, 2015 and installed on the Japanese Experiment Module - Exposed Facility of the ISS. The on-orbit commissioning phase was successfully completed in the first days of October 2015.
Calibration and test of the instrument took place at the CERN-SPS during five campaigns between 2010 and 2015 with beams of electrons, protons and relativistic ions~\cite{akaike2015, bigo, niita}.

\section{Data analysis}
The following analyses are based on flight data (FD)  collected in 2922 days of CALET operation from November 1, 2015 to October 31, 2023 aboard the International Space Station (ISS). The total observation live time for the high-energy (HE) shower trigger is $T\sim 5.5 \times 10^{4}\,h$, corresponding to 86.0\% of the total observation time. Periods when the fild-of-view (FOV) was partially obscured by ISS mobiles structures were excluded from the analysis. Each channel of CHD, IMC and TASC is individually calibrated
on orbit using a dedicated trigger mode which selects penetrating proton and helium particles \cite{niita,CALET2017}.
Raw signals are  corrected  for    gain  differences  among  the  channels,  position and temperature dependence,  temporal gain variations as well as non-uniformity  in  light  output. After calibration, a track is reconstructed for each event with an associated estimate of its charge and energy.

The particle's direction and entrance point are reconstructed and fit by a tracking algorithm based on a combinatorial Kalman filter fed with the coordinates provided by the scintillating fibers of the IMC. It identifies the incident track in the presence of background hits generated by secondary radiation backscattered from the TASC~\cite{paolo2017}.
The angular resolution is $\sim{0.08}^\circ$  for Ti, Cr and Fe and the spatial resolution for the impact point on the CHD is $\sim$180 $\mu$m.

The particle's charge $ Z $ is reconstructed by measuring the ionization deposits in the CHD. The $ dE/dx $ samples are extracted from the signals of the CHD paddles traversed by the incident particle and properly corrected for their path length.
Either CHD layer provides an independent $ dE/dx $ measurement.
In order to correct for the reduction of the scintillator's light yield due to the quenching effect, a ``halo'' model~\cite{GSI} has been used to fit the FD samples of each nuclear species as a function of $ Z^2 $. 
The resulting curves are then used to reconstruct a charge value in either layer ($Z_{\rm CHDX}$, $Z_{\rm CHDY}$) on an event-by-event basis~\cite{CALET-CO}.

The presence of an increasing amount of backscatters from the TASC at higher energy generates additional energy deposits in the CHD that add up to the primary particle ionization signal and may induce an incorrect charge identification.
This effect causes a systematic displacement of the CHDX-CHDY charge peaks to higher values (up to 0.8 $e$ in charge units) with respect to the nominal charge position. Therefore it is necessary to restore the nuclei peak positions to their nominal values  by an energy dependent charge correction applied separately to the FD and the simulated (MC) data. Thanks to the large statistics collected during the first 8 years of operation, it was possible to correct for this effect in the iron and sub-iron region up to a  deposited energy in the TASC  ($E_{\rm TASC}$) of 40 TeV.
A charge distribution obtained by averaging $Z_{\rm CHDX}$ and $Z_{\rm CHDY}$  is shown 
in Fig.~\ref{fig:CHDAVE_100gev}.
\begin{figure} [!htb]
	\centering
	\includegraphics[width=\hsize]{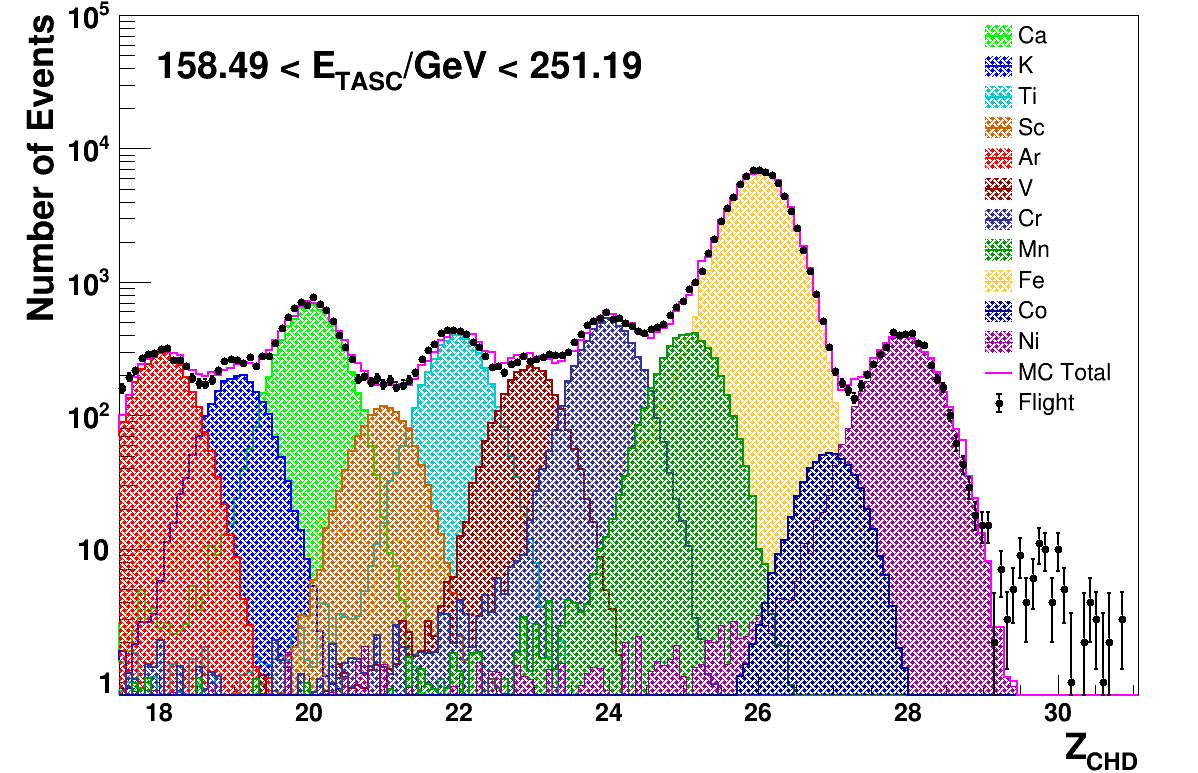}
	\caption{\scriptsize Charge distributions from the combined charge measurement of the two CHD layers in the elemental region between Ar and Zn. 
		Events are selected with $158.49 < E_{\rm TASC} < 251.19$  GeV. Flight data (black dots) are compared with Monte Carlo samples comprising Ar, K, Ca, Sc, Ti, V, Cr, Mn, Fe, Co and Ni. In Fig.~S1 of the SM~\cite{PRL-SM} an enlarged version of this figure is shown. \label{fig:CHDAVE_100gev}}
\end{figure}\noindent
The CHD charge resolution $\sigma_Z$ for iron is $\sim0.35\, e$.

For each event, the shower energy 
  is calculated as
the sum of the energy deposits of all TASC logs, after merging the gain ranges of each channel~\cite{CALET2017}.

The quenching effect in the TASC logs has been included in the MC simulation by extracting a quenching
correction function from the flight data. Further details on this can be found in the SM of Ref.~\cite{CALET-IRON2021}.

Only events with a well-fitted primary track, crossing the detector from CHD top to TASC bottom are used for these analyses.
The  geometrical factor of the fiducial region is  $ S\Omega \sim510\,$cm$^{2}$sr.
 
The  onboard HE trigger based on the coincidence of the summed dynode signals of the last four IMC layers with the TASCX1 layer
has an efficiency close to 100\% for elements heavier than oxygen
as the threshold is far below their signal amplitude  at minimum ionization. Therefore, an off-line trigger confirmation, as required for the analysis of lighter elements \cite{CALET-PROTON,CALET-CO} is not necessary for these analyses. However, we require  a deposited energy larger (by 2 standard deviations) than at minimum ionization  
in at least one of the first four layers of the TASC.
This is adopted to select the shower events which  interact    in the upper part of the TASC (shower event cut).

Particles undergoing a charge-changing nuclear interaction in the upper part of the instrument 
are removed by  applying a selection based on their track width (TW)  and by requiring that the difference between the charges from either layer of the CHD is less than $1.0\,e $  ($1.5\,e $ for iron). 
TW is defined as the difference, normalized to the particle charge, between the total energy deposited in the $\pm$3  fibers nearest to the fiber crossed by the reconstructed track and the sum of the signals of the two fibers adjacent to it.
Examples of TW distributions are shown in Fig. S3 of SM~\cite{PRL-SM}.
Iron events are selected within a wider ellipse than the one used in our previous publication \cite{CALET-IRON2021}: it is centered at Z~=~26, with 2~$ \sigma_x $ and 2~$ \sigma_y $ wide semiaxes for $ Z_{\mathrm{CHDX}} $ and $ Z_{\mathrm{CHDY}}$, respectively,  rotated clockwise by 45 degrees as shown in the cross plot of the CHDY vs CHDX charge in Fig.~S2 of the SM~\cite{PRL-SM}.
Titanium and chromium events, instead, are selected in  CHDY vs CHDX plane  within a circle centered at Z~=~22 and Z~=~24, with a radius $R = 0.55\,e$ and $R = 0.45\,e$, respectively.
Event selections are identical for flight data and MC.

The FOV of CALET ($< 45^\circ $)  is partially shielded at large zenith angle by fixed structures of the ISS. Moreover, moving structures (e.g., solar panels and robotic arms) can cross the FOV for short periods of time during ISS operations. CR interactions in these structures can create secondary nuclei that may induce a contamination of the flux measurement.
Inside the fiducial geometrical factor, 8\% of the final candidate events have reconstructed trajectories pointing to moving structures and they are discarded in the analyses.
Applying the aforementioned cuts, $ 7.4 \times10^{3} $ Ti, $ 6.1 \times10^{3}$ Cr and $ 1.2 \times10^{5} $ Fe candidate events are selected.

An iterative unfolding Bayesian method~\cite{Ago} is applied to correct the distributions  of the total energy deposited in the TASC ($ E_{TASC} $ ) for
significant bin-to-bin migration effects (due to the limited
energy resolution) and infer the primary particle energy.
The response matrix used in the unfolding procedure is derived
using MC simulations after applying the same selection
procedure of the FD and it is shown in Fig.~S6 of the SM~\cite{PRL-SM}. 

The energy spectrum is obtained from the unfolded energy distribution as follows:
\begin{equation}
\Phi(E) = \frac{N(E)}{\Delta E\;  \varepsilon(E) \;  S\Omega \;  T }
\label{eq_flux}
\end{equation}
\begin{equation}
N(E) = U \left[N_{obs}(E_{\rm TASC}) - N_{bg}(E_{\rm TASC}) \right]
\end{equation}
where $\Delta E$ denotes the energy bin width,
$E$ is the geometric mean of the lower and upper bounds of the bin~\cite{Maurino}, 
$N(E)$ the bin content of the unfolded distribution,
$\varepsilon (E)$ the total selection efficiency (Fig.~S4 of the SM~\cite{PRL-SM}), $S \Omega$ is the geometrical factor, $T$ the live time,
$U()$ the unfolding procedure operator,
$N_{obs}(E_{\rm TASC})$ the bin content of the observed energy distribution (including background),
and $N_{bg}(E_{\rm TASC})$ the background events in the same bin.
Background contamination from different nuclear species misidentified as Ti (Cr, Fe) is shown in Fig.~S5
of the SM~\cite{PRL-SM}. 
The contamination fraction $N_{bg}/N_{obs} $ for Fe is very low in the whole energy range ($<2\%$). Instead, the  contaminations for Ti and Cr are larger, though limited to   $ < 11\% $ in the $ E_{TASC} $ energy range between $ 100 $  and $ 4000$ GeV.

\section{Systematic Uncertainties}
The total systematic error  was computed as the
sum in quadrature of all the sources of systematics in each energy bin, both energy-dependent and energy-independent. The energy dependence (in GeV/\textit{n}) of systematic uncertainties is shown in Fig.~S8 of SM~\cite{PRL-SM} for Ti, Cr and Fe.
 The sources of energy-dependent systematic error include (i)  charge identification, (ii) energy scale correction, (iii)  unfolding procedure, (iv)  shower event selection,  (v)  background subtraction and (vi)  track width selection.
The systematic error related to the charge identification for Cr and Ti (Fe) was
studied by varying the values of the semi-minor and major axes of the charge cut up to 30\% (50\%).
The result was a flux variation for Cr and Ti of few \% below 100 GeV/\textit{n}, increasing within $-15\%$ to 10\% at 200  GeV/\textit{n}.
 The iron flux variation ranges from $-3\%$ to 3\%, depending on the energy bin. The uncertainty ($\pm$2\%) on the energy scale from the
beam test calibration affects the absolute normalization of the Cr, Ti and Fe spectra by $\pm$3.5\% but not their shape.
The uncertainty due to the unfolding procedure was
evaluated by using different response matrices computed by varying the spectral index of the generation spectrum of MC simulations. The resulting fluxes did not show appreciable variations.
The contribution due to the shower event cut, which rejects non interacting particles, becomes significant only at low energies (4-5\% below 30 GeV/\textit{n}).
The systematic error related to the background contamination for Cr and Ti  is assessed by varying the contamination level up to $\pm$ 3$ \sigma $ of the statistical error. For Fe, due to lower background, the contamination level is varied up to  $\pm50\%$. The result was
 Cr and Ti fluxes variation around few \% below 100 GeV/\textit{n}, increasing
to 10\% at 200 GeV/\textit{n}; the iron flux normalization remains stable within 1\% in the whole energy range.
The uncertainty on the track width selection leads to a maximum error of 5\% in the Ti flux above
200 GeV/\textit{n} and $\sim 2\%$ below.
Additional energy-independent systematic uncertainties affecting the fluxes normalization include live time (3.4\%) and long-term stability ($ < $ 2\%).

\section{Results}
The energy spectra of Ti, Cr and Fe and their flux
ratio measured with CALET are shown in Fig.~\ref{fig:FlussoTi} and Fig.~\ref{fig:FlussoCr}, respectively; the
corresponding data tables including statistical and systematic errors are reported in SM~\cite{PRL-SM}.

\begin{figure}[h!] 
	\centering

	\includegraphics[width=1\linewidth]{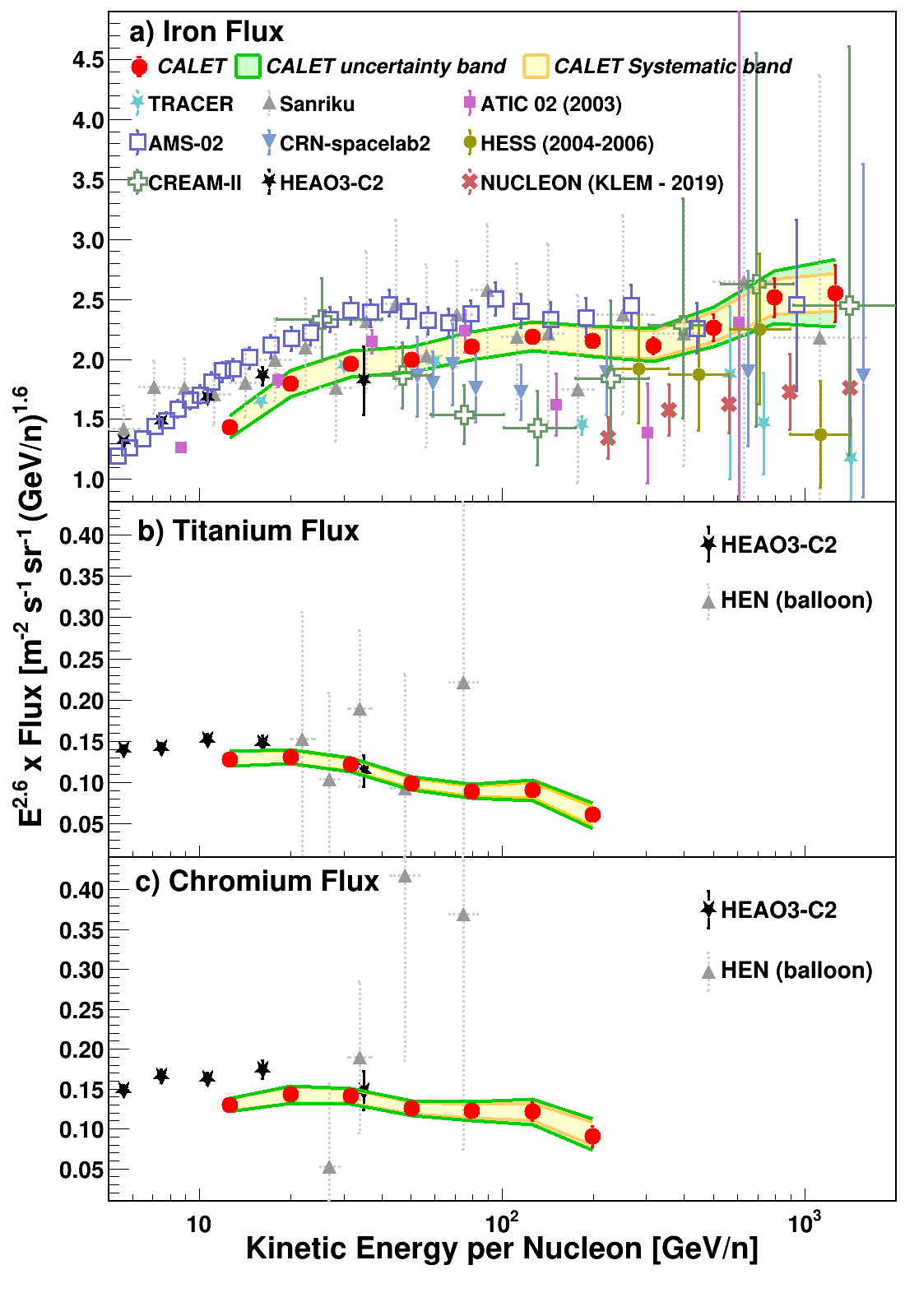} 
	\caption{\scriptsize{CALET (a) Fe, (b) Ti, (c) Cr flux (multiplied by
			$E^{2.6}$), as a function of kinetic
			energy per nucleon.  The error bars of the CALET data (red filled circles) represent the statistical uncertainty only, the yellow band  the quadrature sum of systematic errors, while the green band  the quadrature sum of statistical and systematic errors. Also plotted are other direct measurements~\cite{NUCLEON2019,HEAO,Minagawa,ATIC2, TRACER2008,CRN,AMS-Fe,HESS, CREAM2, HEN}. An enlarged
			version of the figure is shown in Fig. S9   of SM~\cite{PRL-SM}.}}
	\label{fig:FlussoTi}
\end{figure}

\begin{figure}[h!] 
	\centering
	
	\includegraphics[width=1\linewidth]{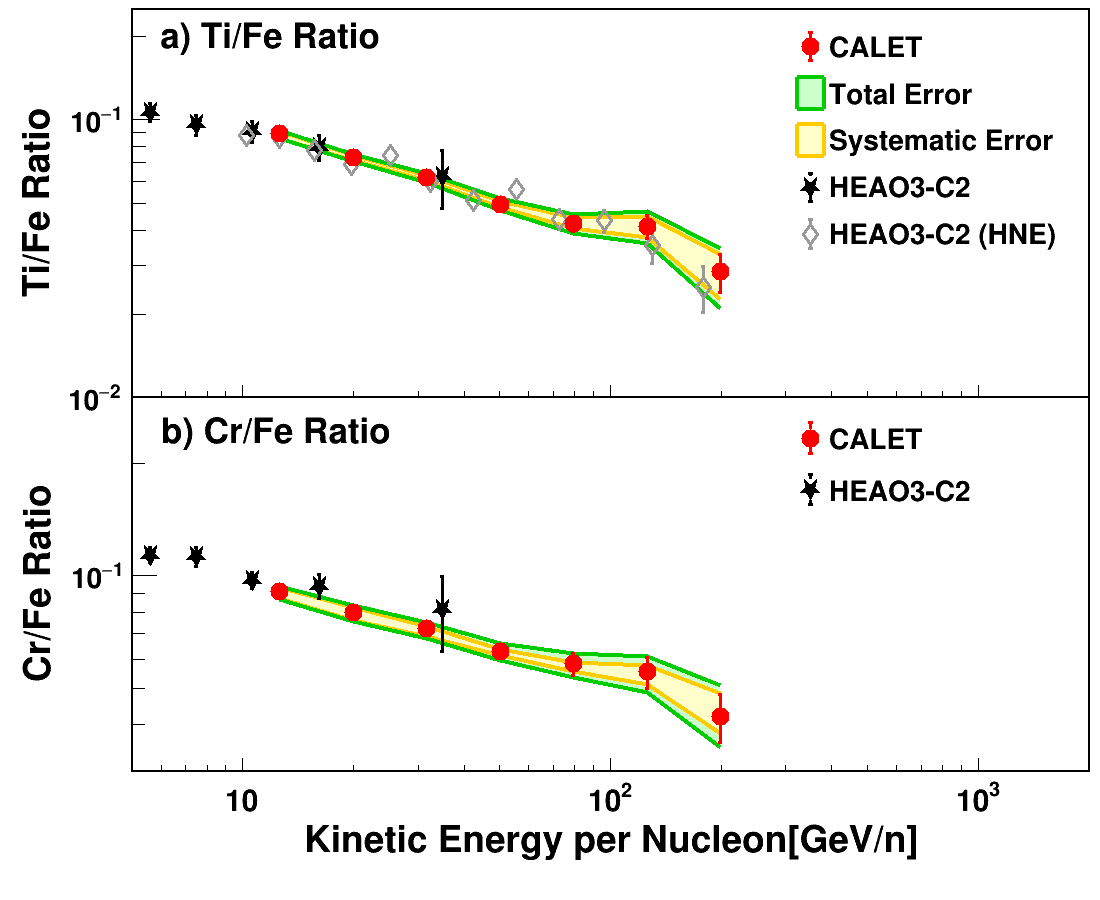}
	\caption{\scriptsize{CALET (a) Ti/Fe and (b) Cr/Fe ratio as a function of kinetic energy per nucleon.  The error bars of the CALET data (red filled circles) represent the statistical uncertainty only, the yellow band  the quadrature sum of systematic errors, while the green band  the quadrature sum of statistical and systematic errors. Also plotted are the HEAO3-C2~\cite{HEAO} and HNE~\cite{HNE_ICRC}  direct measurements.}}
	\label{fig:FlussoCr}
\end{figure}
CALET Cr and Ti spectra are compared with  results from HEAO3-C2~\cite{HEAO} and HEN~\cite{HEN} experiments. The CALET Fe spectrum
turns out to be consistent with most of the previous measurements within the uncertainty error band, both in spectral shape and normalization with the exception of the AMS-02 \cite{AMS-Fe} iron flux measurement which has a very similar shape, but differs in the absolute
normalization  by $ \sim $ 20\%  \cite{CALET-CO, CALET-IRON2021}. Instrumental effects that may cause this tension are under investigation.  The Fe spectrum shown here is based
on a factor of 2 larger data set than in our previous publication and on an improved charge calibration: it is consistent with our earlier result up to 400 GeV/\textit{n}, but differs from it at higher energy, suggesting a possible hardening of the spectrum above a few hundred GeV/\textit{n}.

Figure~\ref{fig:FlussoCr} (a) and (b) show the Ti/Fe and the Cr/Fe ratio, respectively. Also shown are the results from HEAO3-C2 \cite{HEAO} and HNE~\cite{HNE_ICRC} experiments. The two sets of data are in good agreement in the energy interval where they overlap, [10-30] GeV/\textit{n}.
\begin{figure}[h] 
	\centering
	
	\includegraphics[width=1\linewidth]{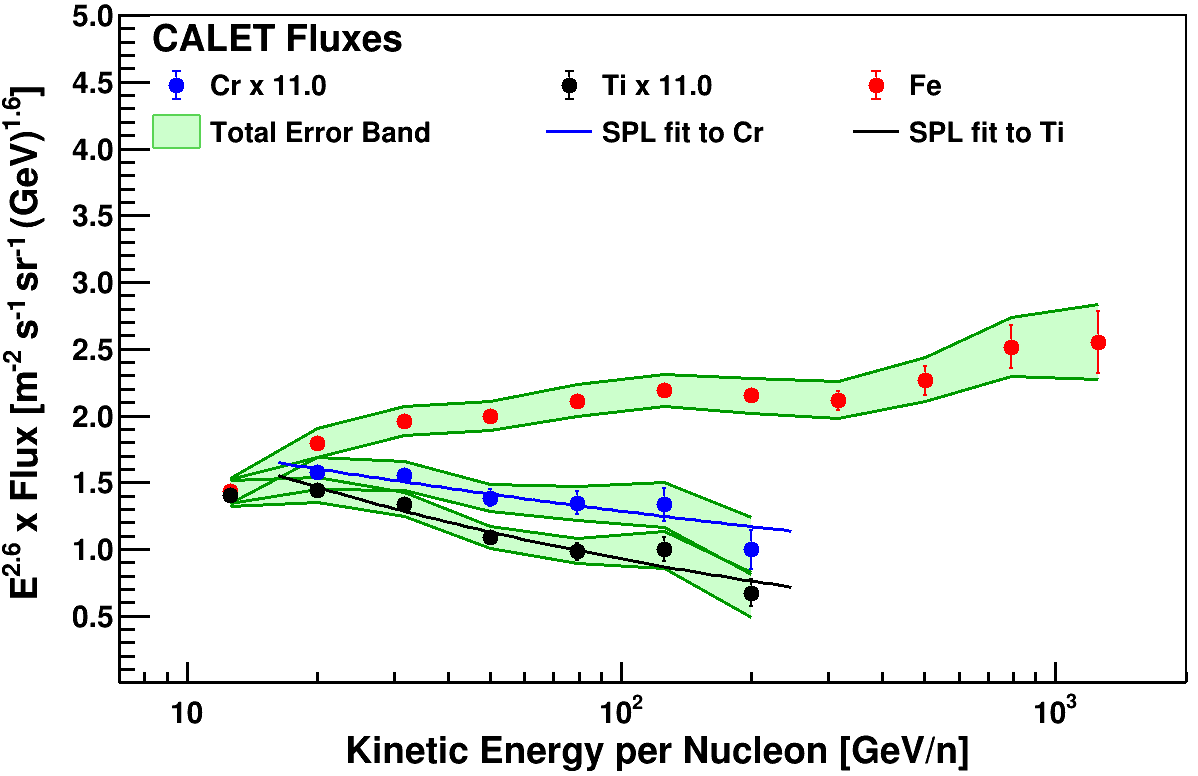}
	\caption{\scriptsize{CALET Cr (blue) and Ti (black) energy spectra are fitted with a SPL function in the energy range [16-250] GeV/\textit{n}; CALET Fe spectrum (red) is  reported for comparison. The Ti and Cr spectra are multiplied by a factor 11 in order to compare them with the iron spectrum.
		Here, the green bands indicate the total error.}}
	\label{fig:FitFlussi}
\end{figure}
Figure~\ref{fig:FitFlussi} shows the fits to CALET Ti and  Cr data with a Single Power-Law function (SPL)   $\Phi(E)= C  \left (\frac{E}{GeV/n}\right) ^\gamma$ where C is a normalization factor, $\gamma$ the spectral index.  The fit is performed from  16  GeV/\textit{n} to 250  GeV/\textit{n} both for Cr and Ti. The fit gives $  \gamma_{\mathrm{Ti}}$ = -2.88 $\pm$ 0.06 (stat.~$\oplus$ syst.)  with $  (\chi^{2} /d.o.f.)_{\mathrm{Ti}} =1.8/4 $ for Ti and  $\gamma_{\mathrm{Cr}} = -2.74\pm 0.06$ (stat.~$\oplus$ syst.) with $  (\chi^{2} /d.o.f.)_{\mathrm{Cr}} =1.1/4 $ for Cr.

An indication of iron spectral hardening is suggested in Fig.~\ref{fig:Finestra_Scorrevole}, where the spectral index is plotted as a function of kinetic energy per nucleon.  We used the iron flux with 10 bins/decade (see Fig.~S7 on SM \cite{PRL-SM}) to calculate the spectral index $\gamma$ above 100 GeV/\textit{n} by  a fit of $d[log (\phi)] /d[log (E)] $ within a sliding window centred on each energy bin and including the $\pm$ 2 neighbouring bins.

\begin{figure}[h] 
	\centering

	\includegraphics[width=1\linewidth]{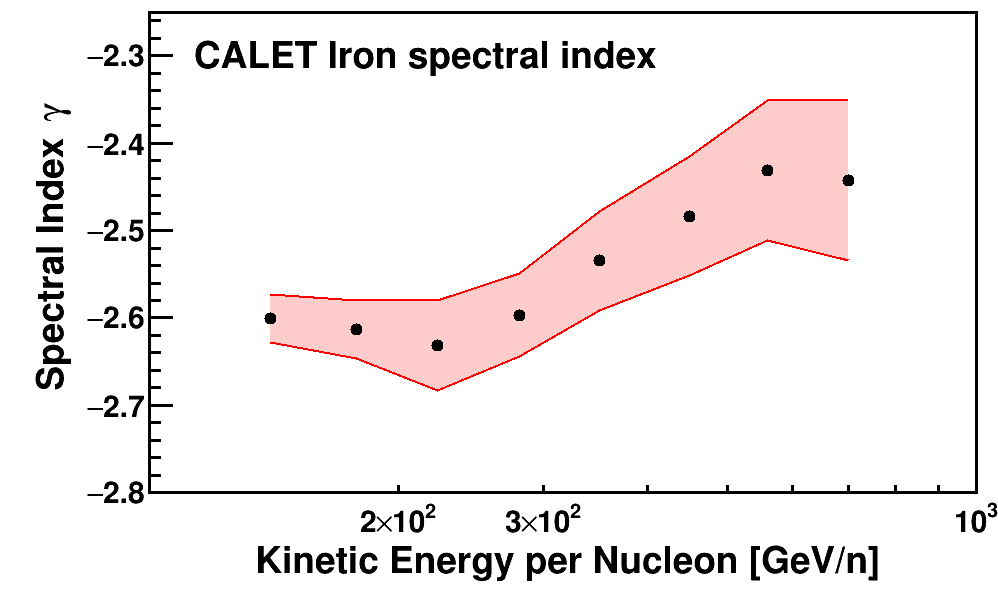}
	\caption{\scriptsize{Energy dependence of the iron spectral index calculated with CALET data within a sliding energy window centred on each energy bin and including $\pm$ 2 neighbouring bins. The shaded area is representative of the uncertainty due to
			purely statistical errors. }}
	\label{fig:Finestra_Scorrevole}
\end{figure}

 \begin{figure}
 	\centering
 
 	\includegraphics[width=1.1\linewidth]{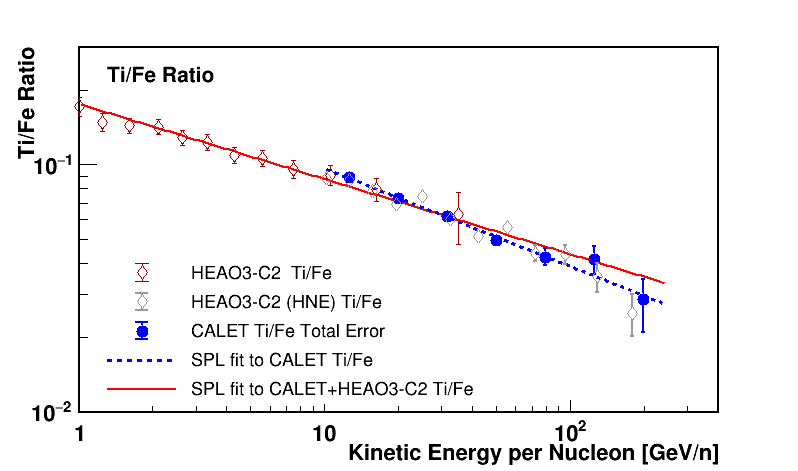}
 	\caption{\scriptsize{Combined fit of CALET and HEAO3-C2~\cite{HEAO} Ti/Fe ratio with a SPL function (red line) in the energy range [1-250] GeV/\textit{n}. Also shown is the fit to the  CALET data alone using  a SPL function (blue dashed line) in the energy range [10-250] GeV/\textit{n}. The error bars of the CALET data represent the total error. The results of HNE experiments~\cite{HNE_ICRC} (magenta diamonds) are  shown for comparison.}}
 	\label{fig:FitRapportiTi}
 \end{figure}

\begin{figure} 
	\centering
	
	\includegraphics[width=1.1\linewidth]{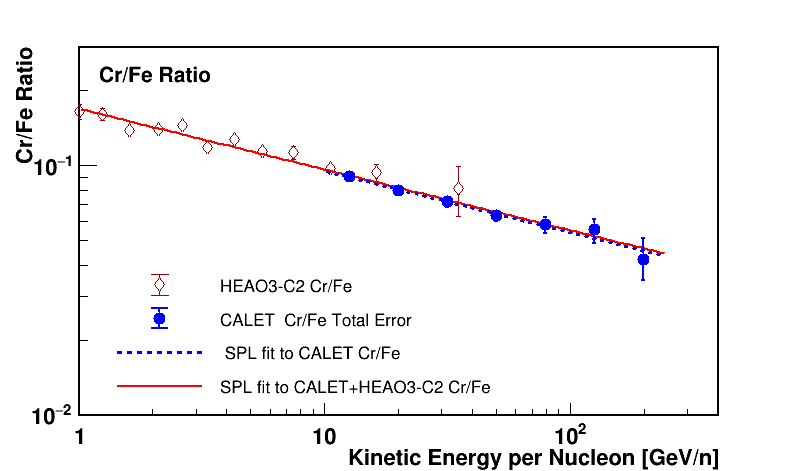}
	\caption{\scriptsize{Combined fit of CALET and HEAO3-C2~\cite{HEAO} Cr/Fe ratio with a SPL function (red line) in the energy range [1-250] GeV/\textit{n}. Also shown is the fit to the CALET data alone using  a SPL function (blue dashed line) in the energy range [10-250] GeV/\textit{n}. The error bars of the CALET data represent the total error.}}
	\label{fig:FitRapportiCr}
\end{figure}

   Figures \ref{fig:FitRapportiTi} and \ref{fig:FitRapportiCr}  show the fits to CALET Ti/Fe and Cr/Fe ratio in the energy range [10-250] GeV/\textit{n} with a SPL fit (blue dashed lines). The best fit spectral indices for these two ratios are $\gamma_{\mathrm{Ti/Fe}}=-0.39\, \pm\, 0.03$ and $\gamma_{\mathrm{Cr/Fe}}=-0.24\, \pm\, 0.03$ with a $\chi ^2/d.o.f. $  of 1.9/5 and 0.8/5, respectively. Also shown in Fig.  \ref{fig:FitRapportiTi} and \ref{fig:FitRapportiCr}  are the fits obtained by combining  the data  from both CALET and HEAO3-C2 experiments in the energy range [1-250] GeV/\textit{n} using with a SPL (red lines): in this case the fit gives $\gamma_{\mathrm{Ti/Fe}}=-0.30 \pm 0.01$ with a $(\chi ^2/d.o.f.)_{\mathrm{Ti/Fe}}=15/17 $  and $\gamma_{\mathrm{Cr/Fe}}=-0.24 \pm 0.01$ with a $(\chi ^2/d.o.f.)_{\mathrm{Cr/Fe}}=17/17 $. The dependence of these indices on the atomic number Z and a comparison with the result obtained by fitting HEAO3-C2 with HNE data  are shown in Fig. \ref{fig:Slope}.
   The variation of the index with Z, as reported in \cite{HNE,HNE_ICRC}, is found to be consistent with the combined HEAO3-C2 and CALET data on Ti and Cr flux ratios to Fe.
  \begin{figure} 
  	\centering
  	\includegraphics[width=1.1\linewidth]{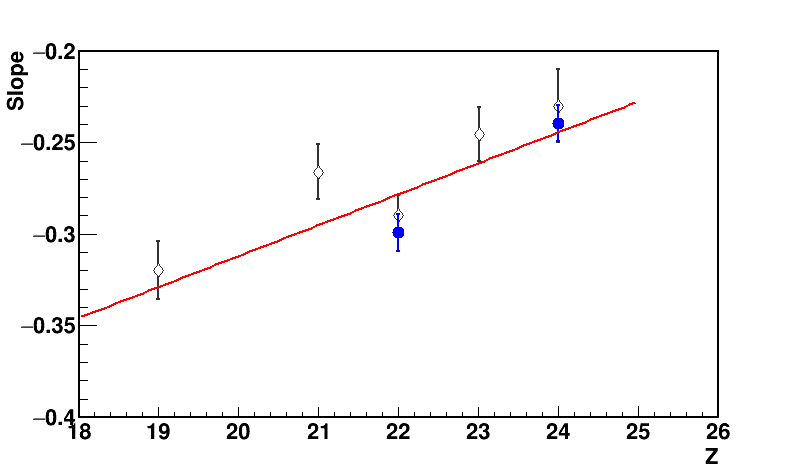}
  	\caption{\scriptsize{
  			Dependence on atomic number Z of the spectral indices of flux ratios to iron. The combined HEAO3-C2 and CALET data on Ti/Fe and Cr/Fe (blue filled circled) are compared with the combined HEAO3-C2 and HNE data (black diamonds)} \cite{HNE,HNE_ICRC}. The red line is the result of a linear fit to the seven  plotted points.}
  	\label{fig:Slope}
  \end{figure}

\section{Conclusion}
In this letter, we present a measurement of the energy spectra of sub-iron elements chromium and titanium from 10 GeV/n to 250 GeV/n, as well as their ratio to iron, with a significantly enhanced precision compared to the  existing measurements.  Our results suggest that the titanium and chromium fluxes, as well as their ratio to iron, are consistent with a SPL within the range of our measurements. Beyond this limit, we will try to extend the observations to draw a significant conclusion regarding a potential deviation from a single power law. The Cr/Fe and Ti/Fe ratios, along with their spectral indices, align with the findings of the HNE and HEAO3-C2 experiments. 
The improvement in statistics with respect to our previous publication \cite{CALET-IRON2021} and a more refined assessment of calibrations and systematics errors have allowed a better determination of the iron spectral shape at high energy, suggesting the possibility of a deviation from a single power law at an energy  exceeding  200 GeV/\textit{n}.

\section{Acknowledgments}
\begin{acknowledgments}
We gratefully acknowledge JAXA's contributions to the development of CALET and to the operations aboard the JEM-EF on the International Space Station.
We also wish to express our sincere gratitude to Agenzia Spaziale Italiana (ASI) and NASA for their support of the CALET project.
This work was supported in part by JSPS Grant-in-Aid for Scientific Research (S) Number 26220708, 19H05608  and 24H00025, 
JSPS Grant-in-Aid for Scientific Research (B) Number 24K00665,  and by the
MEXT-Supported Program for the Strategic Research Foundation at Private Universities (2011-2015)
(No.S1101021) at Waseda University.
The CALET effort in Italy is supported by ASI under agreement 2013-018-R.0 and its amendments.
The CALET effort in the United States is supported by NASA through Grants No. NNX16AB99G, No. NNX16AC02G, and No. NNH14ZDA001N-APRA-0075. We thank Prof. G. Morlino for theoretical discussions on the iron spectral shape.
\end{acknowledgments}
\providecommand{\noopsort}[1]{}\providecommand{\singleletter}[1]{#1}%

\clearpage
\widetext

\setcounter{equation}{0}
\setcounter{figure}{0}
\setcounter{table}{0}
\setcounter{page}{1}
\makeatletter
\renewcommand{\theequation}{S\arabic{equation}}
\renewcommand{\thefigure}{S\arabic{figure}}
\renewcommand{\bibnumfmt}[1]{[S#1]}
\renewcommand{\citenumfont}[1]{S#1}
\begin{center}
	\textbf{\large Precision spectral measurements of  Chromium and Titanium from 10 to 250 GeV/\textit{n}  and sub-Iron to Iron ratio with the Calorimetric Electron Telescope on the ISS \\
		\vspace*{0.5cm}
		SUPPLEMENTAL MATERIAL}	\\
	\vspace*{0.2cm}
	(CALET collaboration) 
\end{center}
\vspace*{1cm}
Supplemental material relative to ``Precision spectral measurements of  Chromium and Titanium from 10 to 250 GeV/\textit{n}  and sub-Iron to Iron ratio with the Calorimetric Electron Telescope (CALET) on the ISS''
\vspace*{1cm}

\clearpage
\section{Additional information on the analysis}

$\bf{Charge \, measurement. \,}$ 
The particle's charge $ Z $ is reconstructed from the ionization deposits in the CHD paddles traversed by the incident particle. Either CHD layer provides an independent $ dE/dx $ measurement which is corrected for the track path length and for the quenching effect in the scintillator's light yield  as a function of $ Z^2 $~\cite{GSI_SM, CALET-CO_SM}. 
For high-energy showers, the charge peaks are corrected for a systematic shift to higher values  with respect to the nominal charge positions, due to the large amount of shower particle tracks backscattered from the TASC whose signals add up to the primary
particle ionization signal.\\
In Fig.~\ref{fig:Z_CHD} the resulting distributions of the reconstructed charge ($Z_{\mathrm{CHD}}$) combining $Z_{\mathrm{CHDX}}$ and
$Z_{\mathrm{CHDY}}$ from flight data (FD) are compared with Monte Carlo (MC) simulations from EPICS.\\
Iron events are selected within an ellipse  centered at $Z = 26$, with 2~$ \sigma_x $ and 2~$ \sigma_y $ wide semi-axes for $ Z_{\mathrm{CHDX}} $ and $ Z_{\mathrm{CHDY}}$, respectively, and rotated clockwise by 45 degrees.
Titanium and chromium events, instead, are selected in  CHDY vs CHDX plane  within a circle centred at $Z = 22$ and $Z = 24 $, with a radius $R = 0.55\,e$ and $R = 0.45\,e$, respectively. 
Charge correlation between $Z_{\mathrm{CHDX}}$ and $Z_{\mathrm{CHDY}}$ in a sample of FD selected 
applying the consistency cut  $|Z_{\mathrm{CHDX}}-Z_{\mathrm{CHDY}}|<1.5$  is shown in Fig.~\ref{fig:CHDX_vs_CHDY} for Ti, Cr and Fe. \\

\noindent
$\bf{Track\, width}.$ The track width (TW) cut allows particles interacting in the initial part of the instrument to be discarded. In fact, the TW distribution of interacting events at the top of the instrument is wider than that of penetrating nuclei, due to the
angular spread of secondary particles produced in the interaction and their lower specific ionization compared to that of the primary particle (see  the SM of \cite{CALET2022BtoC_SM}).
As in each IMC layer neighboring fibers with an energy deposit $>$0.3 MIPs are clustered around the fibers with largest
signals, we define the total track width for heavy nuclei as follows
\[ TW = \frac{1}{6}\sum\limits_{L=1}^6 TW_{L} \] where the sum is limited to the first eight IMC layers from the top excluding the two layers with maximum and minimum energy deposit. The track width in each IMC layer L, $TW_{L}$, is defined as  \[ TW_{L}=\frac{\sum\limits_{j=m-3}^{j=m+3}E_{L,j}-\sum\limits_{j=m-1}^{j=m+1}E_{L,j}} {Z_{CHD}^{2}} \]  where $E_{L,j}$ is the energy deposit in the fiber \textit{j} of the layer \textit{L}, \textit{m} is the index of the fiber with the maximum signal in the cluster crossed by the primary particle track, and the numerator represents the difference between the total
energy deposited in the 7 central fibers of the cluster and the cluster core, encompassing 3 fibers.
$Z_{\mathrm{CHD}}$ represents the charge reconstructed by combining the two layers of the CHD.\\
The TW distribution for a sample of Cr events selected in FD using the charge cut described above, is shown in Fig.~\ref{fig:TW_Cr}. Cr events  are compared with the distributions obtained from MC simulations  of Cr applying the same selections as for FD.\\

\noindent
$\bf{Efficiencies}.$ The total efficiency and relative efficiencies (i.e: the efficiency of a given cut normalized to the previously applied cut) were studied extensively over the whole energy range covered by the titanium, chromium and iron fluxes measurements. 
The total selection efficiency obtained with EPICS  are shown for Ti, Cr, and Fe in Fig.~\ref{fig:tot_eff} as a function of total particle kinetic energy per nucleon. The aforementioned efficiencies were validated by comparing distributions relevant to the event selection, and obtained from flight data, with the same distributions generated by EPICS. The comparison is carried out by appropriately fitting the MC and flight data distributions and comparing the parameters obtained. Any discrepancies between MC and flight data distributions at high energy due to low statistics are taken into account in the
systematics. The latter are limited within  $\pm$5\% for iron over the entire energy range and within $\pm$15\% for Cr and Ti, as shown  in Fig.~\ref{fig:sys_all}. \\

\noindent
$\bf{Interactions \, in\, the \, instrument}.$ The amount of instrument material above the CHD is very small and well known. The largest significant contribution is limited to a 2 mm thick Al cover placed on top of the CHD. This amounts to $\sim2.2\%$ radiation length and $5\times10^{-3}\lambda_{\mathrm{I}} $. The fraction of charge-changing interactions in the Al cover is estimated to be $<$ 1\%. The material description in the MC is very accurate and derived directly from the CAD model.  As CALET is sitting on the JEM external platform of the ISS, no extra material external to CALET is normally present within the acceptance adopted for the flux measurement. 
However, occasional obstructions caused by the ISS robotic arm operations may temporarily affect the FOV. It was checked that the removal of  those rare periods from the data results in a negligible difference in shape and normalization of the flux~($ < $1\%). 
\newline
\indent
MC simulations were used to evaluate the  survival probability after traversing both layers of the CHD and the material above (see SM of \cite{CALET-IRON2021_SM}). The total loss ($\sim 10\%$) of  events interacting in the upper part of the instrument was taken into account in the total efficiency and its uncertainty included in the systematic error.\\

\noindent
$\bf{Background \, contamination.}$ 
A template fit with all MC simulated species from calcium to iron is performed to obtain the correct normalisation with the FD distribution. Once the FD and MC distributions are matched, the background contamination is estimated by counting how many events of each contaminant species fall within the charge selection of Ti (Cr, Fe).
The resulting background contamination of chromium and titanium from neighbouring elements is approximately 11\%, primarily due to iron and, secondarily, to manganese for Cr, and vanadium for Ti. The contamination for iron is negligible throughout the considered energy range and in any case less than 2\%.
The distributions of Ti, Cr and Fe candidates and their contaminants are shown in Fig.~\ref{fig:dnde} as a function of the TASC energy.
The total contamination  was  subtracted bin-by-bin from the selected sample as explained in the main body of the paper. To evaluate the systematic error related to the background for Cr and Ti, the contamination level was varied by $\pm\,3$$\,\sigma$. This resulted  into less than   $ \pm\, 50\% $ variation of the contamination with  a flux
change around few \% below 100 GeV/\textit{n}, increasing to
10\% at 200 GeV/\textit{n}. The iron flux, instead,  exhibits a minimal variation of
1\% in the whole energy range even when the contamination level  varies by $ \pm\, 50\% $.\\

\noindent
$\bf{Calorimetric \, energy, \, bin \, size\,,and \, unfolding.}$
The energy response of the TASC was investigated via  MC simulations and compared with the  measurements of the total particle energy \textit{versus} beam momentum carried out at CERN.  During the beam test campaign of CALET conducted in 2015 at SPS, an extracted primary beam of $^{40}$Ar (150 GeV$ /c/n $) was directed towards an internal target producing beam fragments that were guided towards the instrument along a magnetic beamline.  The latter provided an accurate selection  of their rigidity and A/Z ratio. The relation between the observed TASC energy and the primary energy was measured for several nuclei up to the highest available energy (6~TeV total particle energy in the case of $^{40}$Ar). After an offline rejection of a very small amount of beam contaminants from the data, the shape of the TASC total energy was found to be consistent with a Gaussian distribution (\cite{akaike2015_SM}). 
\newline
\indent
The correlation matrix used for the unfolding was derived from the MC EPICS simulations applying the same selection criteria as in the FD analysis. The normalized unfolding matrices obtained from EPICS  are shown in Fig.~\ref{fig:UFmatrix}  for Ti, Cr and Fe where the color scale indicates the probability for a nucleus candidate with a given primary  energy, of depositing energy in different intervals of $ \mathrm{E_{TASC} }$.  

In order to investigate the effect of the binning, a series of tests were carried out using different binning configurations. The results demonstrated that the smearing matrices are similar, and the final fluxes exhibit almost identical behaviour. For illustrative purposes, the difference between the iron flux with 5 and that with 10 bins per decade is presented in Fig.~\ref{fig:Feflux_with_10_binnings}.
\\

\noindent
$\bf{Energy \, dependent \, systematic \, errors.}$
A breakdown of energy dependent systematic errors stemming from several sources (as explained in the main body of the paper) and including selection cuts, charge identification, energy scale correction, energy unfolding, shower event shape and track width is shown in Fig.~\ref{fig:sys_all} as a function of kinetic energy per nucleon for titanium, chromium and iron. \\

\noindent
$\bf{Difference \, in\, the\, iron\, flux\, with\, respect\, to\, the\, previous\, analysis.}$
The difference in the flux with respect to the  analysis reported in \cite{CALET-IRON2021_SM}  has been extensively investigated. The increase in statistics in the current analysis (twice the previous one) made it possible to re-evaluate the function describing the position of the peaks of the charge distributions in the CHDY as a function of the energy deposited in the TASC. This improved the energy dependence of charge selection in the region between 600 GeV/\textit{n} and 2 TeV/\textit{n}. In fact, in this range, the systematic
uncertainty related to the charge cut, obtained by varying the semi-axes of the ellipse by  $\pm$15\%, was about  $\pm$10\%, in the previous analysis. In the new analysis, the systematic uncertainty related to the charge cut, (obtained by varying the
semi-axes of the ellipse by  $\pm$50\%) is reduced to  $\pm$3\%. In addition, statistical errors that ranged from  $\pm$10\% to  $\pm$30\% in the previous analysis have been reduced to no more than 10\% in the current one. \\

\noindent
$\bf{Titanium \, chromium \ and  \ iron \, flux \, normalization \,and \, spectral \, shape.}$
The CALET Ti, Cr and Fe fluxes and a compilation of available data are shown in Fig.~\ref{fig:FluxSM}, providing an enlarged version of Fig.~2 in the main body of the~paper. 
\newline
\indent\
\begin{figure}[htb!]  \centering
	\subfigure[]{\includegraphics[width=0.8\hsize]{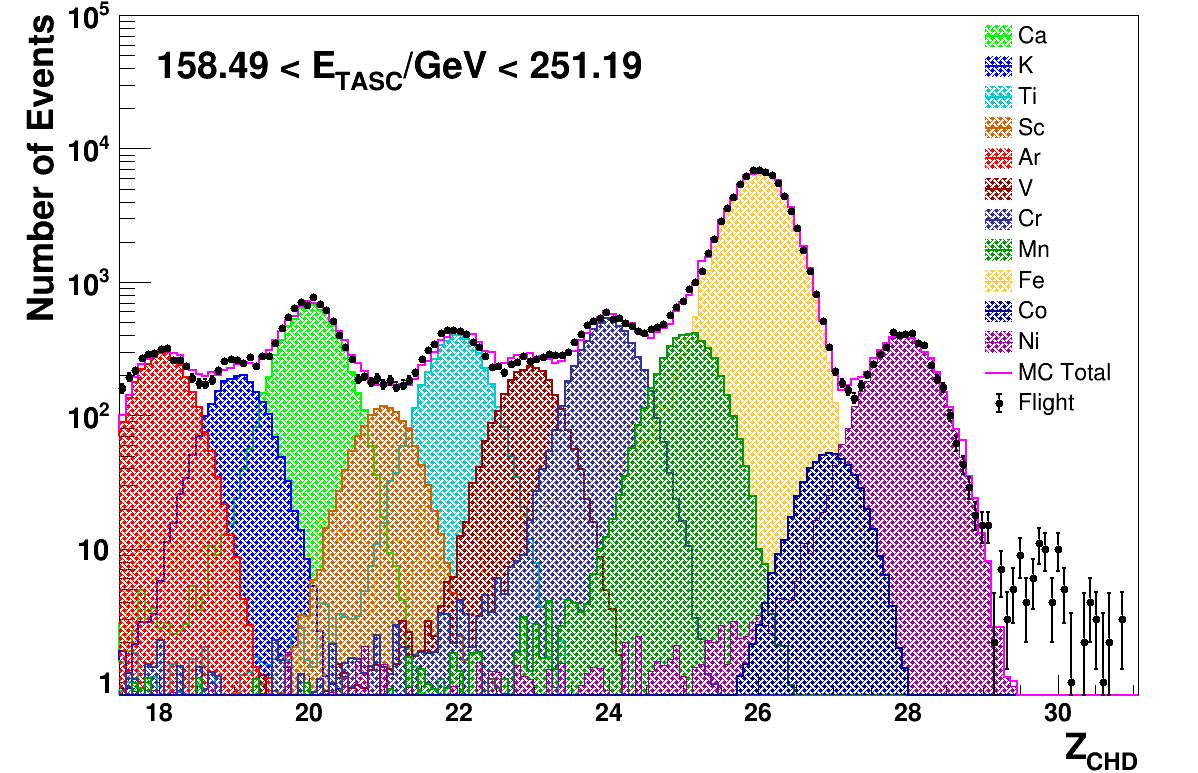}}
	\subfigure[]{\includegraphics[width=0.8\hsize]{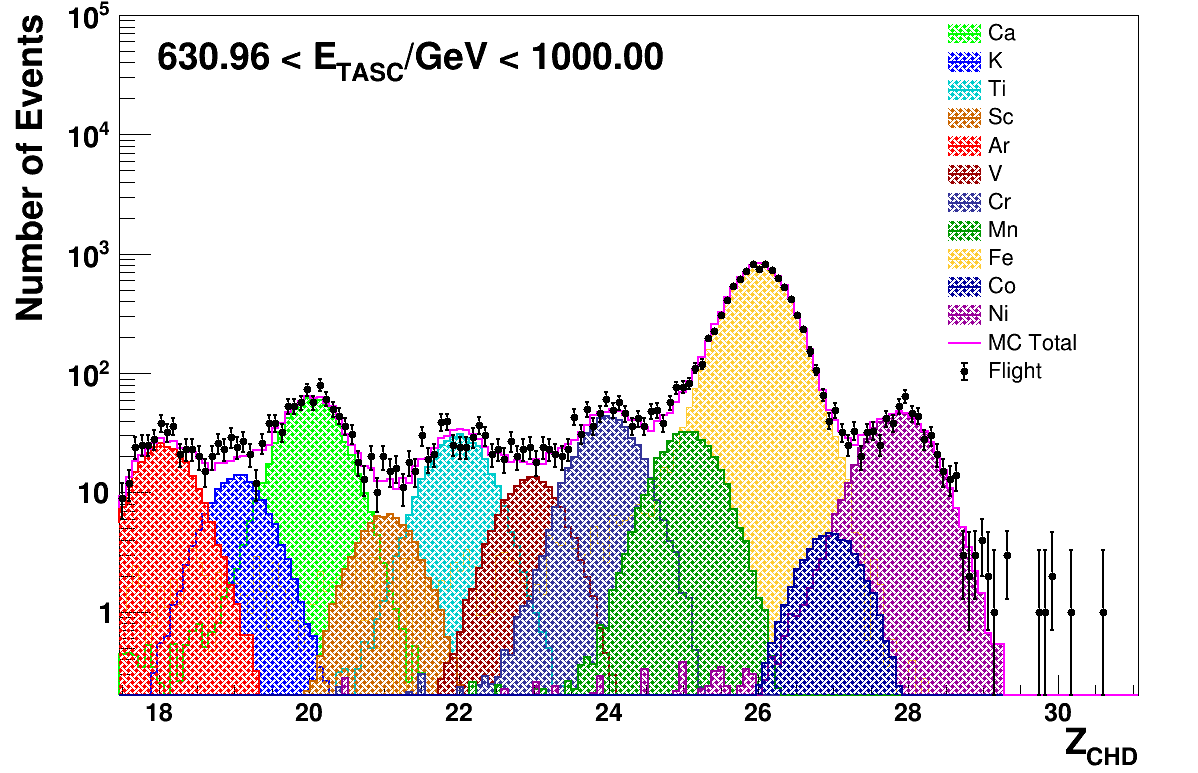}}
	\caption{\scriptsize Charge distributions from the combined charge measurement of the two CHD layers in the elemental region between Ar and Zn. 
		Events are selected with $158.49 < E_{\rm TASC} < 251.19$  GeV in (a) and $630.96 < E_{\rm TASC} < 1000$ GeV in (b). Flight data, represented by black dots, are compared with Monte Carlo samples including argon, potassium, calcium, scandium, titanium, vanadium, chromium, manganese, iron, cobalt and nickel.}
	\label{fig:Z_CHD}
\end{figure}\noindent
\begin{figure} [!htb]
	\centering
	\includegraphics[width=0.46\hsize]{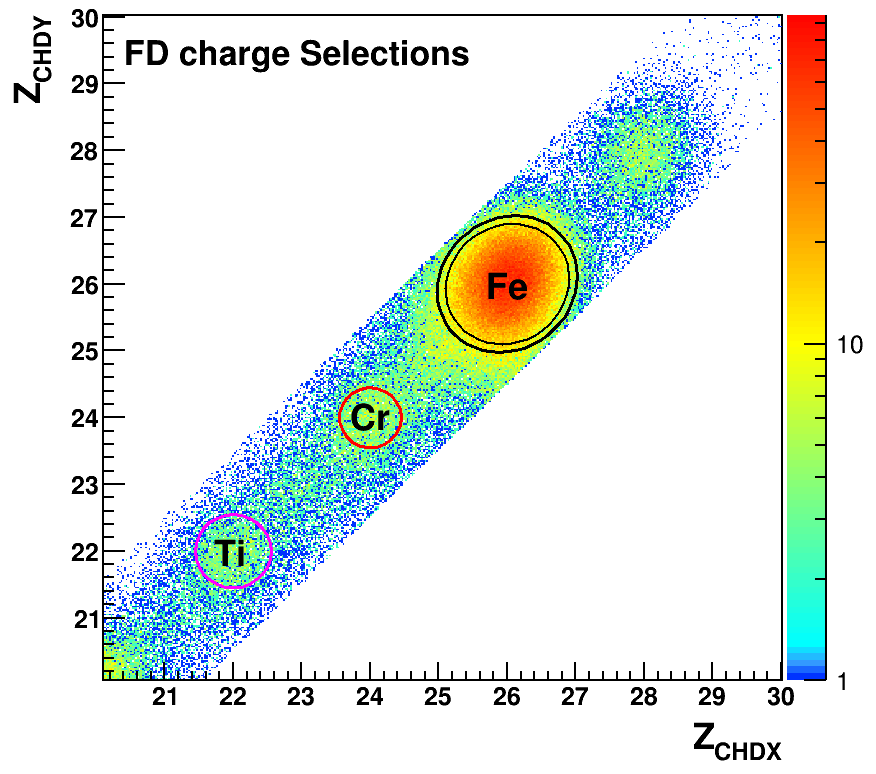}
	\caption{\scriptsize Crossplot of $ Z_{\rm CHDY} $ vs. $ Z_{\rm CHDX} $ reconstructed charges. Events are selected with $E_{\rm TASC}> 63.1$  GeV. The maximum and the minimum elliptical selection depending on the energy for the iron candidates are indicated by the black ellipses in the figure. The charge selection for the titanium and chromium  candidates is indicated by the magenta and red   circles, respectively. \label{fig:CHDX_vs_CHDY}}
\end{figure}\noindent
\begin{figure} [!htb]
	\centering
	\includegraphics[width=0.8\hsize]{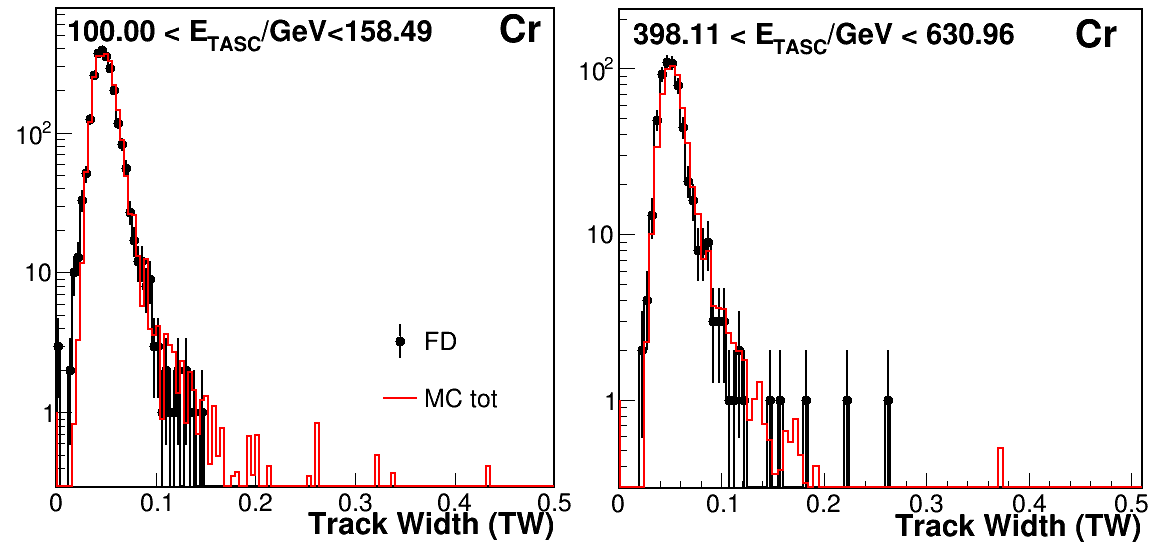}
	\caption{\scriptsize  Track width (TW) distribution in two different intervals of $ E_{\mathrm{TASC}} $. The black dots represent a sample
		of Cr events selected in FD by means of the CHD. The red line represents the MC distribution of the sum of chromium and background nuclei (Ti, V, Mn, Fe). MC distributions  are obtained with the same selections used for FD. \label{fig:TW_Cr}}
\end{figure}\noindent


\begin{figure}[!htb] \centering
	\includegraphics[width=0.5\hsize]{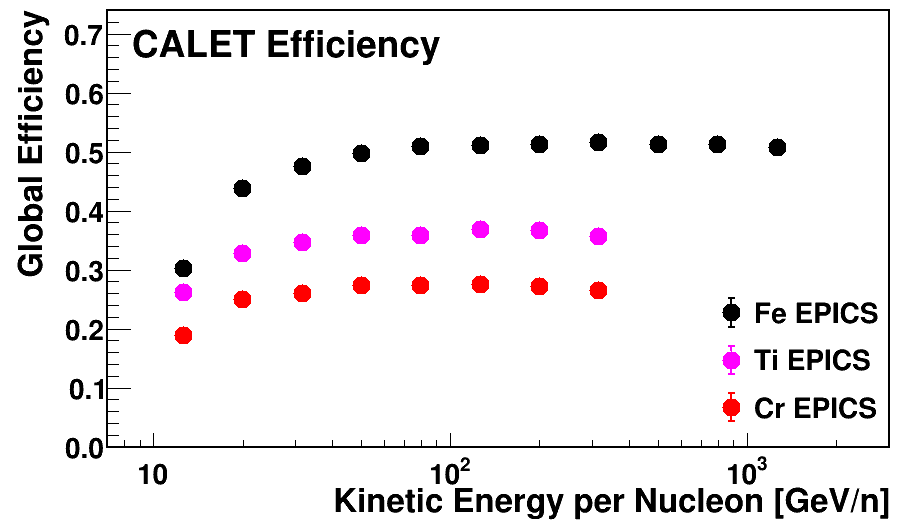}
	\caption{\scriptsize{Total selection efficiency for titanium, chromium and iron events as estimated with EPICS simulations.  The lower efficiency of chromium versus titanium is due to the narrower charge cut. 
	}}
	\label{fig:tot_eff}
\end{figure}

\begin{figure*}[htb] 
	\begin{minipage}[s]{0.49\textwidth}
		\includegraphics[width=\textwidth]{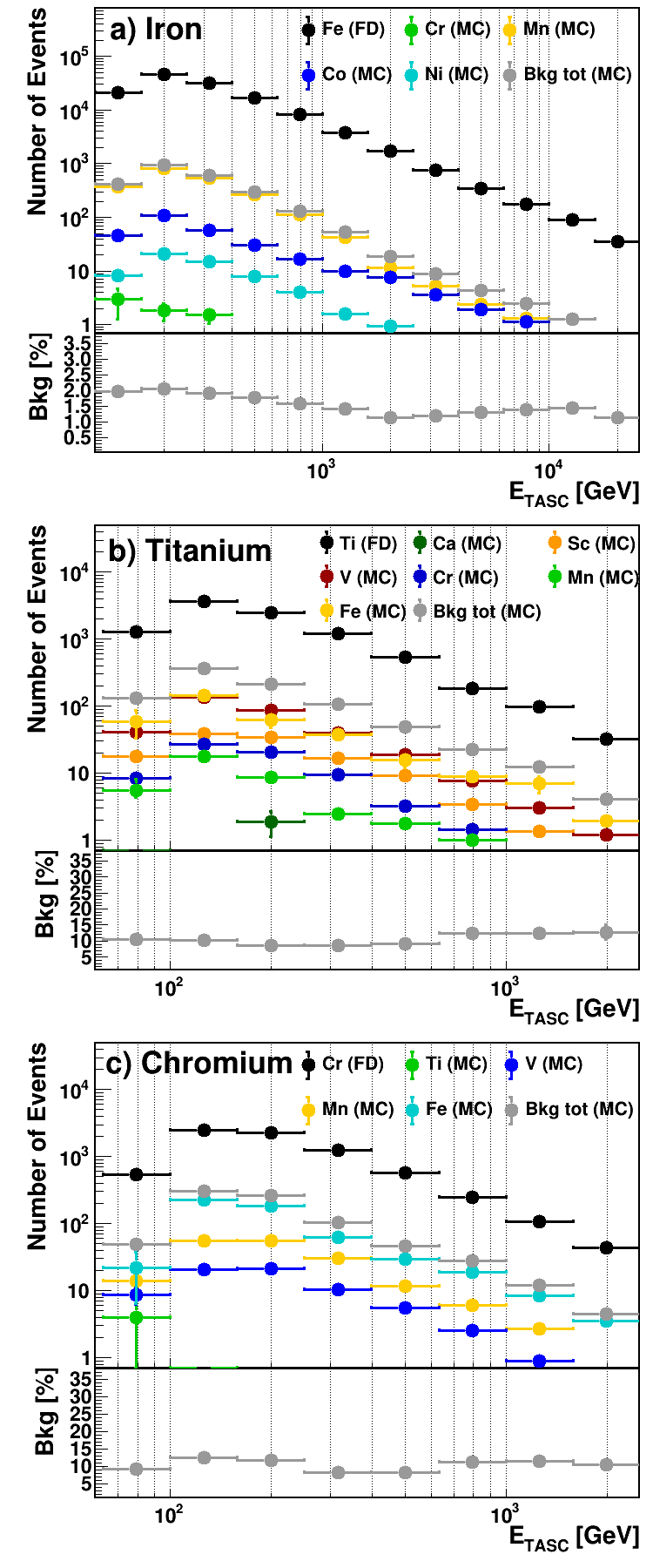}
		\caption{\scriptsize{ Top panels: Differential distributions of the number of events in a given bin of calorimetric energy ($E_{\rm TASC}$ in GeV) for selected (a) iron, (b) titanium and (c) chromium events in flight data (black dots) before the unfolding procedure and with background events from neighbouring nuclei. Bottom panels: contamination from neighbouring nuclei obtained with the MC.}} 
		\label{fig:dnde}
	\end{minipage}
	\hfill 
	\begin{minipage}[s]{0.43\textwidth}
		\includegraphics[width=\textwidth]{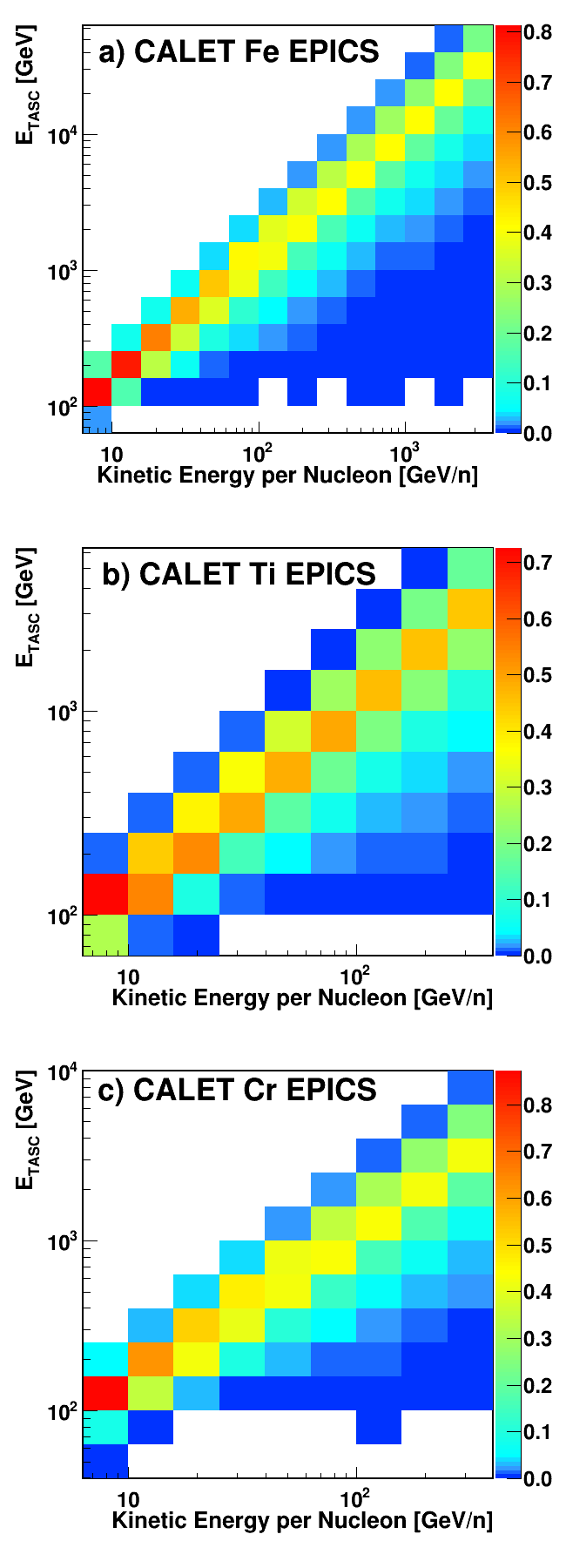}
		\caption{\scriptsize{Response matrix for (a) iron, (b) titanium, and (c) chromium  derived from the MC simulations of the CALET flight model by applying the same selections as for flight data. The array is normalized so that the color scale is associated with the probability that  candidates, in a given bin of particle kinetic energy, may correspond to different intervals of $ E_\mathrm{TASC}$.}}
		\label{fig:UFmatrix}
	\end{minipage}
\end{figure*}
\begin{figure}[!htb] \centering
	\includegraphics[width=0.8\hsize]{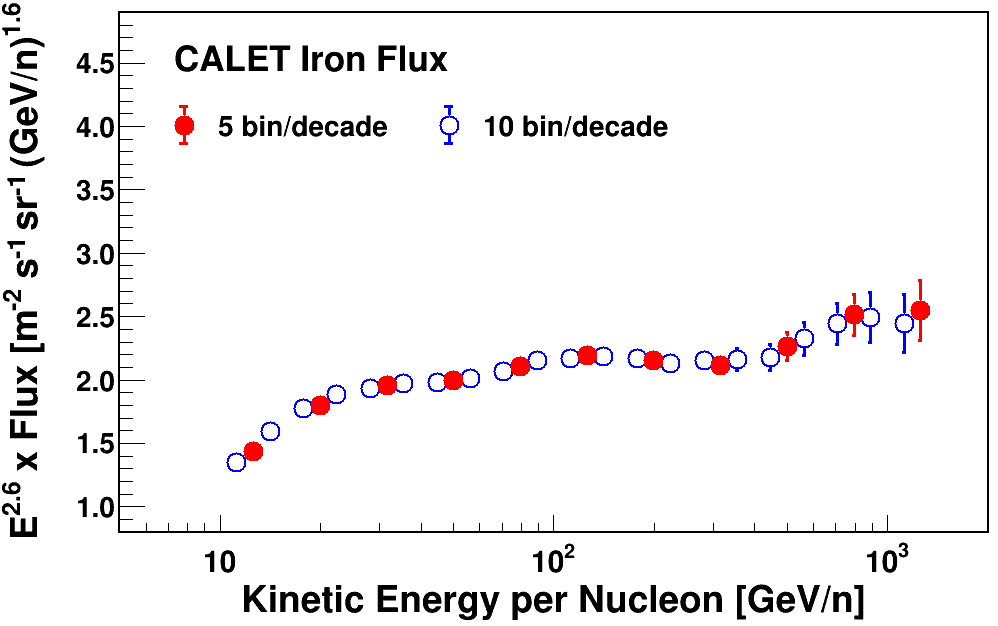}
	\caption{\scriptsize{CALET iron flux with standard binning (red circles) and 10 bins/decade (blue open circles). The vertical error bars are representative of purely statistical errors.
	}}
	\label{fig:Feflux_with_10_binnings}
\end{figure}\noindent

\begin{figure}[htb!]  \centering
	\subfigure[]{\includegraphics[width=0.62\hsize]{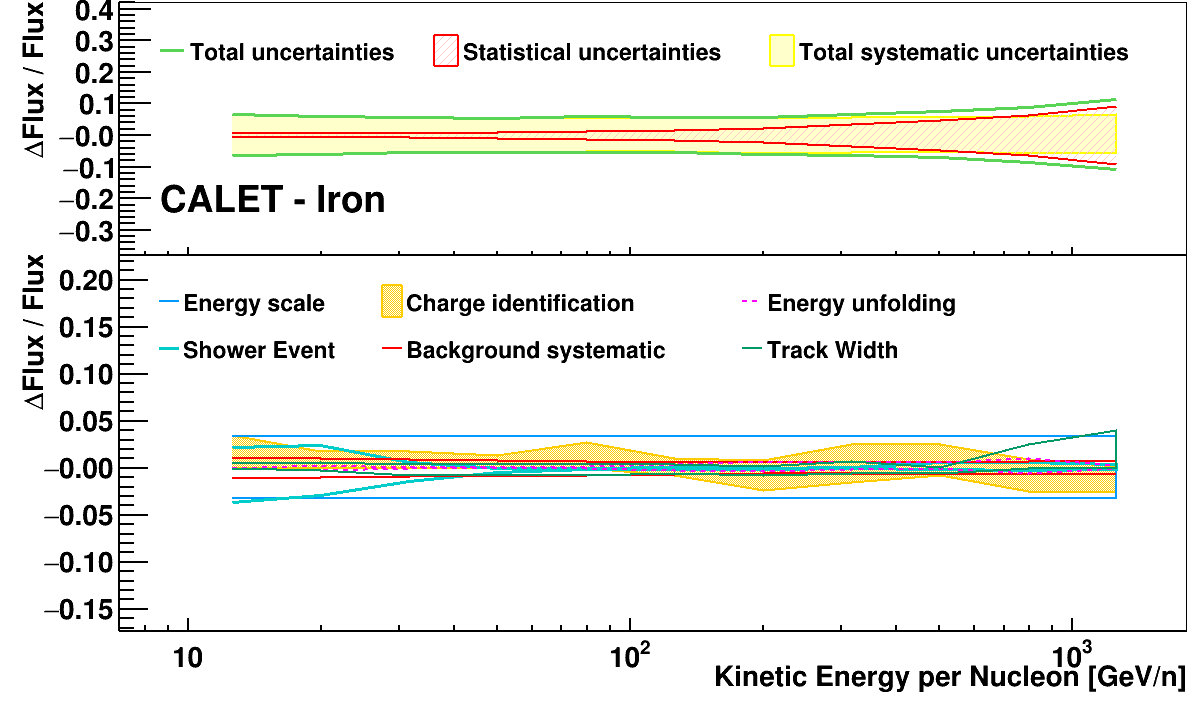}}\\
	\subfigure[]{\includegraphics[width=0.62\hsize]{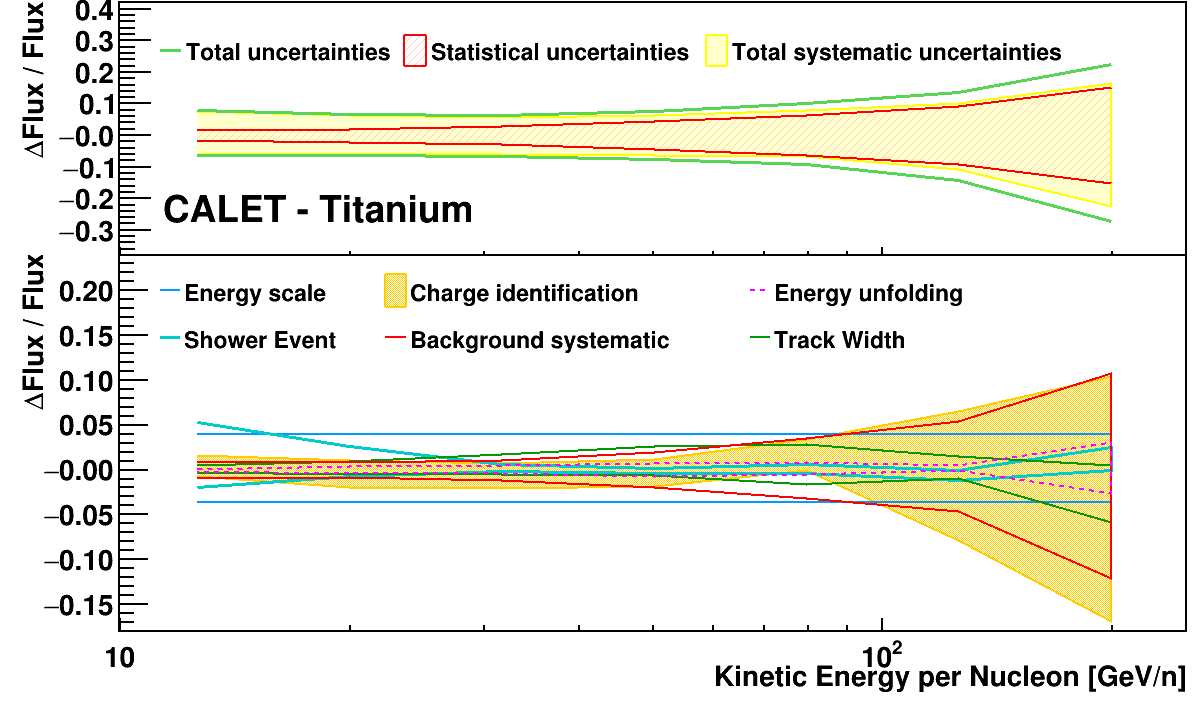}}\\
	\subfigure[]{\includegraphics[width=0.62\hsize]{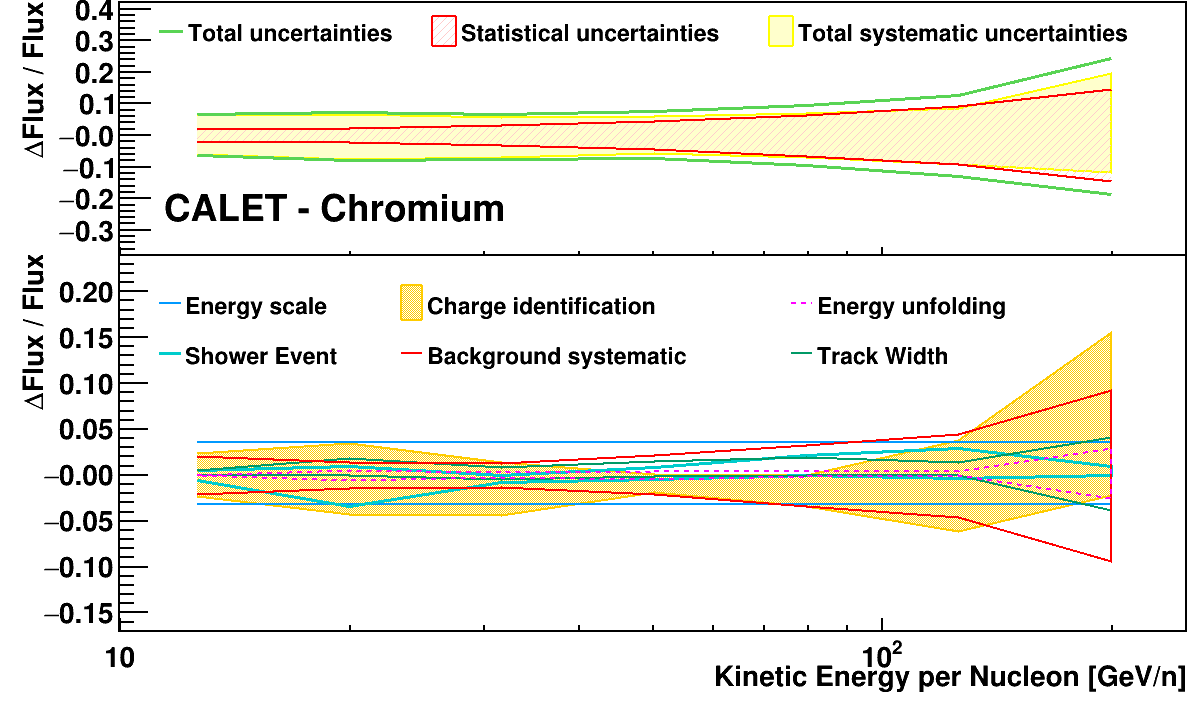}}
	\caption{\scriptsize{Energy dependence (in GeV/\textit{n}) of systematic uncertainties (relative errors) for (a) iron, (b) titanium and (c) chromium.  In the upper panels, the band bounded by the red lines represents the statistical error. The yellow band shows the sum in quadrature of all the sources of systematics including energy independent ones. The green lines represent the sum in quadrature of statistical and total systematic uncertainties. In the bottom panels, a detailed breakdown of systematic energy dependent errors, stemming from charge identification, energy scale correction, energy unfolding,  shower event shape and track width is shown.}}
	\label{fig:sys_all}
\end{figure}\noindent

\begin{figure}[!htb]
	\centering
	\includegraphics[width=0.9\hsize]{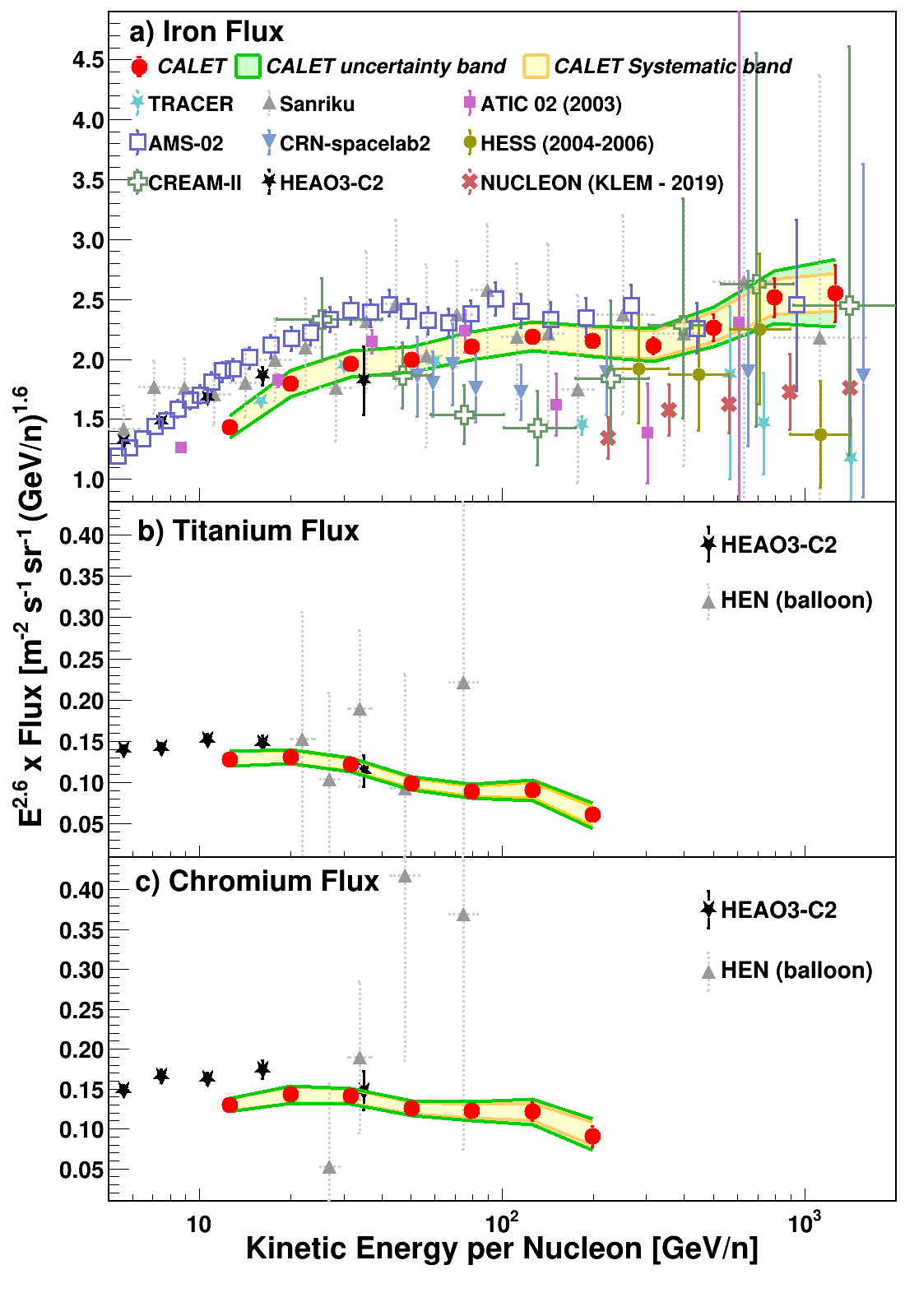}
	\caption{\scriptsize{CALET (a) Fe, (b) Ti, (c) Cr  flux as a function of kinetic energy per nucleon in GeV (with multiplicative factor $E^{2.6}$). The error bars of the CALET data (red filled circles) represent the statistical uncertainty only. The yellow band indicates the quadrature sum of systematic errors, while the green band indicates the quadrature sum of statistical and systematic errors. Also plotted are the data points from NUCLEON ~\cite{NUCLEON2019_SM}, HEAO3-C2~\cite{HEAO_SM}, Balloon (Sanriku)~\cite{Minagawa81_SM}, ATIC~\cite{ATIC2_SM}, TRACER~\cite{TRACER2008_SM}, CRN~\cite{CRN_SM}, AMS-02~\cite{AMS-Fe_SM}, HESS~\cite{HESS_SM}, HEN~\cite{HEN_SM} and CREAM~\cite{CREAM2_SM}. \label{fig:FluxSM}}}
\end{figure}

\clearpage
\renewcommand{\arraystretch}{1.25}
\begin{table*}[!htb]
	\caption{Table of the CALET differential spectrum in kinetic energy per nucleon of cosmic-ray titanium. 
		The first, second, and third error in the flux are representative of the statistical uncertainties, systematic uncertainties in normalization, and energy dependent systematic uncertainties, respectively.
		\label{tab:Cflux}}
	\begin{ruledtabular}
		\begin{tabular}{c c c c}
			Energy Bin [GeV/n] & Flux [m$^{-2}$sr$^{-1}$s$^{-1}$(GeV/n)$^{-1}$]   \\
			\hline
			10.0 -- 15.8 & $( 1.77 \, \pm 0.03 \, \pm 0.07 \, _{- 0.08 }^{+ 0.12}) \times 10^{- 4 }$ \\
			15.8 -- 25.1 & $( 5.47 \, \pm 0.11 \, \pm 0.21 \, _{- 0.25 }^{+ 0.26}) \times 10^{- 5 }$ \\
			25.1 -- 39.8 & $( 1.53 \, \pm 0.04 \, \pm 0.06 \, _{- 0.07 }^{+ 0.07 }) \times 10^{- 5 }$ \\
			39.8 -- 63.1 & $( 3.77 \, \pm 0.16 \, \pm 0.15 \, _{- 0.18 }^{+ 0.19 }) \times 10^{- 6 }$ \\
			63.1 -- 100.0 & $( 1.02 \, \pm 0.07 \, \pm 0.04 \, _{- 0.06 }^{+ 0.07 }) \times 10^{- 6 }$ \\
			100.0 -- 158.5 & $( 3.15 \, \pm 0.29 \, \pm 0.12 \, _{- 0.32 }^{+ 0.30 }) \times 10^{- 7 }$ \\
			158.5 -- 251.2 & $( 6.41 \, \pm 0.98 \, \pm 0.25 \, _{- 1.42 }^{+ 1.02}) \times 10^{- 8 }$ \\
		\end{tabular}
	\end{ruledtabular}
\end{table*}
\renewcommand{\arraystretch}{1.0}

\renewcommand{\arraystretch}{1.25}
\begin{table*}[!htb]
	\caption{Table of the CALET differential spectrum in kinetic energy per nucleon of cosmic-ray chromium. 
		The first, second, and third error in the flux are representative of the statistical uncertainties, systematic uncertainties in normalization, and energy dependent systematic uncertainties, respectively.
		\label{tab:Cflux}}
	\begin{ruledtabular}
		\begin{tabular}{c c c c}
			Energy Bin [GeV/n] & Flux [m$^{-2}$sr$^{-1}$s$^{-1}$(GeV/n)$^{-1}$]   \\
			\hline
			10.0 -- 15.8 & $( 1.80 \, \pm 0.04 \, \pm 0.07 \, _{- 0.08 }^{+ 0.09 }) \times 10^{- 4 }$ \\
			15.8 -- 25.1 & $( 5.97 \, \pm 0.14 \, \pm 0.23 \, _{- 0.39 }^{+ 0.33 }) \times 10^{- 5 }$ \\
			25.1 -- 39.8 & $( 1.78 \, \pm 0.06 \, \pm 0.07 \, _{- 0.10 }^{+ 0.08 }) \times 10^{- 5 }$ \\
			39.8 -- 63.1 & $( 4.79 \, \pm 0.22 \, \pm 0.19 \, _{- 0.21 }^{+ 0.22 }) \times 10^{- 6 }$ \\
			63.1 -- 100.0 & $( 1.41 \, \pm 0.09 \, \pm 0.06 \, _{- 0.08 }^{+ 0.08 }) \times 10^{- 6 }$ \\
			100.0 -- 158.5 & $( 4.21 \, \pm 0.39 \, \pm 0.16 \, _{- 0.35 }^{+ 0.32 }) \times 10^{- 7 }$ \\
			158.5 -- 251.2 & $( 9.49 \, \pm 1.37 \, \pm 0.37 \, _{- 1.06 }^{+ 1.82 }) \times 10^{- 8 }$ \\
		\end{tabular}
	\end{ruledtabular}
\end{table*}
\renewcommand{\arraystretch}{1.0}
\renewcommand{\arraystretch}{1.25}
\begin{table*}[!htb]
	\caption{Table of the CALET differential spectrum in kinetic energy per nucleon of cosmic-ray iron. 
		The first, second, and third error in the flux are representative of the statistical uncertainties, systematic uncertainties in normalization, and energy dependent systematic uncertainties, respectively.
		\label{tab:Cflux}}
	\begin{ruledtabular}
		\begin{tabular}{c c c c}
			Energy Bin [GeV/n] & Flux [m$^{-2}$sr$^{-1}$s$^{-1}$(GeV/n)$^{-1}$]   \\
			\hline
			10.0 -- 15.8 & $( 1.98 \, \pm 0.01 \, \pm 0.08 \, _{- 0.10 }^{+ 0.11 }) \times 10^{- 3 }$ \\
			15.8 -- 25.1 & $( 7.48 \, \pm 0.04 \, \pm 0.29 \, _{- 0.34 }^{+ 0.35 }) \times 10^{- 4 }$ \\
			25.1 -- 39.8 & $( 2.47 \, \pm 0.02 \, \pm 0.10 \, _{- 0.09 }^{+ 0.10 }) \times 10^{- 4 }$ \\
			39.8 -- 63.1 & $( 7.60 \, \pm 0.07 \, \pm 0.30 \, _{- 0.26}^{+ 0.28 }) \times 10^{- 5 }$ \\
			63.1 -- 100.0 & $( 2.42 \, \pm 0.03 \, \pm 0.09 \, _{- 0.08 }^{+ 0.11 }) \times 10^{- 5 }$ \\
			100.0 -- 158.5 & $( 7.60 \, \pm 0.13 \, \pm 0.30 \, _{- 0.26 }^{+ 0.28 }) \times 10^{- 6 }$ \\
			158.5 -- 251.2 & $( 2.26 \, \pm 0.05 \, \pm 0.09 \, _{- 0.09 }^{+ 0.08 }) \times 10^{- 6 }$ \\
			251.2 -- 398.1 & $( 6.69 \, \pm 0.23 \, \pm 0.26 \, _{- 0.25 }^{+ 0.29 }) \times 10^{- 7 }$ \\
			398.1 -- 630.9 & $( 2.16 \, \pm 0.10 \, \pm 0.08 \, _{- 0.07 }^{+ 0.09 }) \times 10^{- 7 }$ \\
			630.9 -- 1000.0 & $( 7.26 \, \pm 0.47 \, \pm 0.28 \, _{- 0.30 }^{+ 0.33 }) \times 10^{- 8 }$ \\
			1000.0 -- 1584.9 & $( 2.22 \, \pm 0.20 \, \pm 0.09 \, _{- 0.09 }^{+ 0.12 }) \times 10^{- 8 }$ \\
		\end{tabular}
	\end{ruledtabular}
\end{table*}
\renewcommand{\arraystretch}{1.0}
\renewcommand{\arraystretch}{1.25}
\begin{table*}[!htb]
	\caption{Table of CALET titanium to iron flux ratio. The first and the second error in the flux are representative of the statistical  uncertainties and systematic uncertainties, respectively.
		\label{tab:Cflux}}
	\begin{ruledtabular}
		\begin{tabular}{c c c c}
			Energy Bin [GeV/n] & Flux [m$^{-2}$sr$^{-1}$s$^{-1}$(GeV/n)$^{-1}$]   \\
			\hline
			10.0 -- 15.8 & $( 8.90 \, \pm 0.15 \, \pm  \, _{- 0.31 }^{+ 0.17 }) \times 10^{- 2 }$ \\
			15.8 -- 25.1 & $( 7.31 \, \pm 0.16 \, \pm  \, _{- 0.16 }^{+ 0.11 }) \times 10^{- 2 }$ \\
			25.1 -- 39.8 & $( 6.21 \, \pm 0.18 \, \pm  \, _{- 0.15 }^{+ 0.08 }) \times 10^{- 2 }$ \\
			39.8 -- 63.1 & $( 4.96 \, \pm 0.22 \, \pm  \, _{- 0.13 }^{+ 0.11 }) \times 10^{- 2 }$ \\
			63.1 -- 100.0 & $( 4.24 \, \pm 0.27 \, \pm  \, _{- 0.18 }^{+ 0.21 }) \times 10^{- 2 }$ \\
			100.0 -- 158.5 & $( 4.15 \, \pm 0.39 \, \pm  \, _{- 0.38 }^{+ 0.35 }) \times 10^{- 2 }$ \\
			158.5 -- 251.2 & $( 2.84 \, \pm 0.44 \, \pm  \, _{- 0.59 }^{+ 0.43 }) \times 10^{- 2 }$ \\
		\end{tabular}
	\end{ruledtabular}
\end{table*}
\renewcommand{\arraystretch}{1.0}
\renewcommand{\arraystretch}{1.25}
\begin{table*}[!htb]
	\caption{Table of CALET chromium to iron flux ratio. The first and the second error in the flux are representative of the statistical  uncertainties and systematic uncertainties, respectively.
		\label{tab:Cflux}}
	\begin{ruledtabular}
		\begin{tabular}{c c c c}
			Energy Bin [GeV/n] & Flux [m$^{-2}$sr$^{-1}$s$^{-1}$(GeV/n)$^{-1}$]   \\
			\hline
			10.0 -- 15.8 & $( 9.07 \, \pm 0.19 \, \pm  \, _{- 0.36 }^{+ 0.23 }) \times 10^{- 2 }$ \\
			15.8 -- 25.1 & $( 7.98 \, \pm 0.19 \, \pm  \, _{- 0.36 }^{+ 0.29 }) \times 10^{- 2 }$ \\
			25.1 -- 39.8 & $( 7.23 \, \pm 0.23 \, \pm  \, _{- 0.33 }^{+ 0.12 }) \times 10^{- 2 }$ \\
			39.8 -- 63.1 & $( 6.30 \, \pm 0.29 \, \pm  \, _{- 0.18 }^{+ 0.05 }) \times 10^{- 2 }$ \\
			63.1 -- 100.0 & $( 5.82 \, \pm 0.38 \, \pm  \, _{- 0.28 }^{+ 0.06 }) \times 10^{- 2 }$ \\
			100.0 -- 158.5 & $( 5.54 \, \pm 0.52 \, \pm  \, _{- 0.42 }^{+ 0.21 }) \times 10^{- 2 }$ \\
			158.5 -- 251.2 & $( 4.21 \, \pm 0.62 \, \pm  \, _{- 0.40 }^{+ 0.65 }) \times 10^{- 2 }$ \\
		\end{tabular}
	\end{ruledtabular}
\end{table*}
\renewcommand{\arraystretch}{1.0}

\newpage
\providecommand{\noopsort}[1]{}\providecommand{\singleletter}[1]{#1}%

\end{document}